\documentclass{aa}  
\usepackage{graphicx}
\usepackage{natbib}
\usepackage{txfonts}
\begin{document}
   \title{Infrared spectroscopy of HCOOH in interstellar ice analogues}

   \subtitle{}

   \author{S.~E. Bisschop \inst{1} \and G.~W. Fuchs \inst{1} \and
A.~C.~A. Boogert \inst{2} \and E.~F. van Dishoeck\inst{1} \and
H. Linnartz\inst{1}}

   \offprints{S.E. Bisschop, bisschop@strw.leidenuniv.nl}

   \institute{Raymond and Beverly Sackler Laboratory for Astrophysics,
Leiden Observatory, Leiden University, P.O. Box 9513, 2300 RA Leiden,
The Netherlands \and {AURA/NOAO-South, Casilla 603, La Serena, Chili}}

   \date{Received; accepted}

 
  \abstract
   {HCOOH is one of the more common species in interstellar ices with abundances of 1--5\% with respect to solid H$_2$O. With the launch of the Spitzer Space Telescope new infrared spectra have become available of interstellar ices in different environments. So far systematic laboratory studies on HCOOH-containing interstellar ice analogues are lacking. }
   {This study aims at characterizing the HCOOH spectral features in astrophysically relevant ice mixtures in order to interpret astronomical data.}
   {The ices are grown under high vacuum conditions and spectra are recorded in transmission using a Fourier transform infrared spectrometer. Pure HCOOH ices deposited at 15~K and 145~K are studied, as well as binary and tertiary mixtures containing H$_2$O, CO, CO$_2$ and CH$_3$OH. The mixture concentrations are varied from 50:50\% to $\sim$10:90\% for HCOOH:H$_2$O. Binary mixtures of HCOOH:X and tertiary mixtures of HCOOH:H$_2$O:X with X = CO, CO$_2$, and CH$_3$OH, are studied for concentrations of $\sim$10:90\% and $\sim$7:67:26\%, respectively.}
   {Pure HCOOH ice spectra show broad bands which split around 120~K due to the conversion of a dimer to a chain-structure. Broad single component bands are found for mixtures with H$_2$O. Additional spectral components are present in mixtures with CO, CO$_2$ and CH$_3$OH. The resulting peak position, full width at half maximum and band strength depend strongly on ice structure, temperature, matrix constituents and the HCOOH concentration. Comparison of the solid HCOOH 5.9, 7.2, and 8.1~$\mu$m features with astronomical data toward the low mass source HH~46 and high mass source W~33A shows that spectra of binary mixtures do not reproduce the observed ice features. However, our tertiary mixtures especially with CH$_3$OH match the astronomical data very well. Thus interstellar HCOOH is most likely present in tertiary or more complex mixtures with H$_2$O, CH$_3$OH and potentially also CO or CO$_2$, providing constraints on its formation.}
   {}

   \keywords{Astrochemistry, Line: profiles, Molecular data, Molecular processes, Methods: laboratory, ISM: molecules, Infrared: ISM}

   \maketitle
%

\section{Introduction}

Formic acid is one of the few molecules in star forming regions that
is detected both in the solid state and in the gas phase \citep[see
e.g.,][]{schutte1999,ikeda2001}. Infrared observations of ices in high
mass Young Stellar Objects (YSOs) with the Infrared Space Observatory
(ISO) contain absorption features that are assigned to HCOOH and that
correspond to the C=O stretch, $\nu_{\rm S}$(C=O), at 5.9~$\mu$m and
the OH and CH bending modes, $\nu_{\rm B}$(OH) and $\nu_{\rm B}$(CH)
at 7.2~$\mu$m \citep{schutte1997,schutte1999}. The $\nu_{\rm S}$(C=O)
band is the strongest spectral feature of HCOOH ice, but suffers from
partial overlap with the OH bending mode of solid H$_2$O at 6~$\mu$m,
$\nu_{\rm B}$(OH). The $\nu_{\rm B}$(OH/CH) band at 7.2~$\mu$m is
significantly weaker than the $\nu_{\rm S}$(C=O) band, but is present
in a relatively clean region and has previously been detected toward
high mass star forming regions \citep{schutte1999}. More recently,
these features have also been observed toward low mass YSOs and
possibly toward background stars \citep[Boogert et al. in
prep.;][]{knez2005}. Typical abundances of HCOOH vary between
individual sources from 1 to 5\% with respect to solid H$_2$O. The aim
of this paper is to study the $\nu_{\rm S}$(C=O), $\nu_{\rm B}$(OH/CH)
and other infrared features of HCOOH ices in different astrophysically
relevant laboratory matrices and to compare the results with
astronomical spectra to infer the chemical environment of HCOOH in
interstellar ices.

In the gas phase HCOOH is detected in envelopes surrounding high and
low mass star forming regions, as well as in shocked regions toward
the galactic center
\citep{dishoeck1995,ikeda2001,bottinelli2007,requena2006,bisschop2007a}. HCOOH
is part of the surface chemistry network leading to the formation of
complex organic molecules seen in hot cores \citep{tielens1997} and is
therefore expected to be present in compact hot regions where it has
evaporated from dust grains. In contrast, interferometric observations
show that gas phase HCOOH emission can be extended beyond the hot core
and is not coincident with other oxygen-bearing species such as
CH$_3$OCHO \citep{liu2002,hollis2003,remijan2006}. The observed gas
phase HCOOH has low rotation temperatures and its abundance is a
factor of 10$^{-4}$ lower than that of solid HCOOH
\citep{bisschop2007a}. It is very likely that gaseous and solid HCOOH
co-exist in the same region. A low level of non-thermal desorption due
to e.g., cosmic ray spot heating is sufficient to explain the observed
gas phase abundances of HCOOH.

Previously, laboratory ice spectra have been studied for pure HCOOH
and mixtures of HCOOH with H$_2$O and/or CH$_3$OH by
\citet{schutte1999} and \citet{hudson1999}. These studies only focused
on specific spectral features. The data clearly showed that the
morphology and composition of the ice matrix strongly influence the
HCOOH infrared spectrum. Recently, experiments have been reported for
the layered ice system HCOOH/H$_2$O in which the interaction of HCOOH
and H$_2$O in the interface is studied \citep[see
e.g.,][]{cyriac2005,bahr2005,souda2006}.

There is a clear need to understand the spectroscopy of the main ice
constituents such as HCOOH, as new observational tools become
available for the study of infrared features of interstellar ices,
such as the Spitzer Space Telescope, the Stratospheric Observatory For
Infrared Astronomy (SOFIA), and the James Webb Space Telescope
(JWST). Solid HCOOH is likely mixed with species such as H$_2$O, CO,
CO$_2$ and CH$_3$OH that are abundant in the ice and have related
formation mechanisms. Infrared ice features of mixtures with these
species have been studied in this paper. Pure HCOOH is measured as
well for comparison. Quantitative and qualitative analysis of the
5.9~$\mu$m, 7.2~$\mu$m and other HCOOH ice features allows for a
better determination of the amount of HCOOH present in interstellar
ices as well as its ice environment.

The paper is organized as follows: Sect.~\ref{exp_method} explains the
experimental method, Sect.~\ref{analysis} the analysis techniques,
Sect.~\ref{res} discusses the results of experiments with pure ices,
as well as mixtures with H$_2$O, CH$_3$OH, CO and CO$_2$, in
Sect.~\ref{bandstrength} the effects of ice matrix and temperature on
the band strength are presented, Sect.~\ref{chemdisc} discusses the
physical aspects of the interactions of HCOOH in the solid state with
other species, Sect.~\ref{astro} presents the astrophysical
implications, and finally Sect.~\ref{summary} summarizes the main
conclusions of this paper.

\section{Experimental procedure}\label{exp_method}

The experiments are performed in a high vacuum (HV) set-up with a base
pressure of $\sim$10$^{-7}$~Torr, that has been previously described
in detail by \citet{gerakines1995}. A CsI window mounted in close
thermal contact with a closed cycle He cryostat is situated at the
center of the chamber. It is used as a substrate holder and can be
cooled down to 14~K. The sample temperature is controlled to better
than $\pm$0.1~K using the cryostat cold finger, a resistive heating
element and a Lakeshore 330 temperature control unit. The system
temperature is monitored by two KP-type thermocouples (0.07\% Au in Fe
versus chromel), one mounted on the substrate face and the second
close to the heater element. Ices are grown {\it in situ} onto the
substrate, by exposing the cold surface to a steady flow of gas that
is introduced into the chamber via an all metal flow control valve
with a modified outlet directed at the substrate center along the
surface normal. The ices are monitored by Fourier Transform Infrared
(FTIR) absorption spectroscopy with 1~cm$^{-1}$ spectral resolution
covering the 4000--500~cm$^{-1}$ range.

To test the effect of dilution of HCOOH (98\% purity, J.~T. Baker) by
H$_2$O (deionized) four mixtures are prepared in a glass vacuum
manifold. Both HCOOH and H$_2$O are further purified by subsequently
freezing and thawing the samples in the vacuum manifold. The mixture
ratios range from the relative abundances derived from observations of
interstellar ices of 1 to 5\% \citep[see e.g.,][]{gibb2004} to more
HCOOH-rich ices with relative HCOOH:H$_2$O concentrations up to
50:50\%. The astrophysical motivation for these higher mixture ratios
is that other species such as CH$_3$OH are present in solid state
environments with nearly equal amounts of H$_2$O and/or CO$_2$
\citep{ehrenfreund1998,ehrenfreund1999,dartois1999}. The different
HCOOH:H$_2$O mixtures that are used here are summarized in
Table~\ref{sample}. A pure H$_2$O sample is prepared for
comparison. Furthermore, some features of the solid CO band can be
explained with mixtures or layers of CO with non-hydrogen-bonding
molecules or with hydrogen-bonding species other than H$_2$O (e.g.,
\citealp{tielens1991}; \citealp{pontoppidan2003}; Fraser et al., in
prep.). Accordingly, we have also studied a set of binary mixtures of
HCOOH:X and tertiary mixtures of HCOOH:H$_2$O:X, where X stands for CO
(99.997\% purity, Praxair), C$^{18}$O$_2$ (97\% purity, Praxair) or CH$_3$OH
(99.9\% purity, Janssen Chimica) are prepared with concentrations of
$\sim$10:90\% and $\sim$7:67:26\%, respectively. These mixtures are
chosen such that the matrix species determine the structure of the
ice. Uncertainties on the concentrations of the ice constituents are
$\sim$10\% due to inaccuracies in the pressure reading. The
C$^{18}$O$_2$ isotopologue is used to distinguish solid C$^{18}$O$_2$
from regular atmospheric gas phase CO$_2$.

\begin{table}
\caption{Overview of ice morphologies, total ice exposure and
deposition time.}\label{sample} \centering
\begin{tabular}{lll|ll}
\hline
\hline
H$_2$O & HCOOH & X & Total exposure & Deposition time\\
  (\%)     & (\%)      & (\%)  & (L)      & (min)\\
\hline
--     & 100   & -- & \phantom{1}900           & 15\\ 
100    & --    & -- & \phantom{1}900           & 15\\
50   & 50  & -- & \phantom{1}900               & 15\\
66   & 34  & -- & \phantom{1}900               & 15\\
80   & 20  & -- & \phantom{1}900               & 15\\
91   & 9   & -- & \phantom{1}900             & 15\\
-- & 11 & 89(CO) & 1800             & 30\\
-- & 9 & 91(C$^{18}$O$_2$) & \phantom{1}900  & 15\\
-- & 10 & 90(CH$_3$OH) & 1800    & 30\\
62  & 8     & 30(CO)       & 1800     & 30\\
67  & 6     & 27(C$^{18}$O$_2$) & 1800 & 30\\
68  & 6     & 26(CH$_3$OH) & 1800     & 30\\
\hline
\end{tabular}
\end{table}

A typical experiment starts with taking a background spectrum at
15~K. Subsequently, the ices are deposited with a flow of
$\sim$1.0$\times$10$^{-5}$~Torr~s$^{-1}$. An exposure of
1.0$\times$10$^{-6}$~Torr~s$^{-1}$ corresponds to 1~Langmuir (L) or
equivalently 1~monolayer~(ML)~s$^{-1}$ assuming that the molecular
surface density is 10$^{15}$~molecules~cm$^{-2}$ and the sticking
probability is 1 at 15~K. The deposition times and total exposures are
summarized in Table~\ref{sample}. After deposition, FTIR spectra are
taken at 15~K, and subsequently every 15~K for binary mixtures of
HCOOH with H$_2$O or CO$_2$ and every 30~K for all other mixtures up
until a temperature of 165~K is reached. At each temperature, the
sample stabilizes for 20~minutes before a spectrum is recorded.

One pure HCOOH experiment is performed where HCOOH is deposited at
145~K. The aim of this experiment is to determine whether the HCOOH
ice structure depends on the deposition temperature and whether
potential phase changes are reversible. Subsequently, the experiment
is cooled down to 15~K and from then on FTIR spectra are taken every
15~K during warm-up.

\section{Data reduction and analysis}\label{analysis}

\begin{table*}
\caption{Frequency ranges used for baseline subtraction.}\label{base}
\begin{center}
\begin{tabular}{ll}
\hline
\hline
Ice composition & Baseline range \\
 & (cm$^{-1}$) \\
\hline
HCOOH \& HCOOH:H$_2$O   & 4000--3800, 1850--1800, 1050--1030, 540--500\\
HCOOH:CH$_3$OH \& HCOOH:H$_2$O:X$^a$ & 4000--3700, 1850--1800, 1580--1500, 950--900, 500--450 \\
HCOOH:CO                  & 4000--3800, 2440--2260, 2000--1800, 1550--1400, 600--500 \\
HCOOH:CO$_2$              & 4000--3900, 2100--1800, 1550--1430, 550--500 \\
\hline
\end{tabular}
\end{center}
$^a$ X = CO, C$^{18}$O$_2$ or CH$_3$OH.
\end{table*}

The spectral range from 4000--500~cm$^{-1}$ is very rich in absorption
features of complex organic species, and HCOOH is no
exception. Consequently special care has to be taken in the data
reduction. Depending on the mixture, different frequency ranges are
used to fit third order polynomial baselines (see
Table~\ref{base}). An additional local baseline is subtracted in the
1800--1100~cm$^{-1}$ frequency range. This is necessary as the HCOOH
$\nu_{\rm S}$(C=O) mode around $\sim$1700~cm$^{-1}$ (5.9~$\mu$m)
overlaps with the $\nu_{\rm B}$(OH) mode of H$_2$O at 1655~cm$^{-1}$
(6.0~$\mu$m). This H$_2$O feature is much weaker than the HCOOH
feature, even when only 10\% of HCOOH is present, but it is
responsible for a wing on the $\nu_{\rm S}$(C=O) band. Finally, a
correction is made on the $\nu_{\rm B}$(OH/CH) feature around
1390~cm$^{-1}$ (7.2~$\mu$m) as it is close to a variable background
feature. Spectra that comprise only the background feature are
subtracted from the sample spectra for each temperature, but due to
its variability, the subtraction does not fully remove the artifact.

The absolute band strengths are difficult to calibrate in our
experiment, because the number of molecules that stick to the
substrate sample is not exactly known. Instead, we focus in this paper
on the relative band strengths that can be accurately measured. The
relative values discussed here can be combined with previously
measured values for the band strengths available from the literature
(summarized in Table~\ref{ab_band}). The absolute band strengths for
the vibrational modes depend on whether HCOOH is in the gas phase or
solid state and consequently the numbers in the table do not perfectly
match.

\begin{table*}
  \caption{Summary values absolute band strengths for HCOOH from the literature.}\label{ab_band}
\begin{tabular}{llll}
\hline
\hline
Ice matrix & $\nu_{\rm S}$(C=O) & $\nu_{\rm S}$(C--O) & $\nu_{\rm B}$(OH/CH) \\
                   & cm~molecule$^{-1}$ & cm~molecule$^{-1}$ & cm~molecules$^{-1}$\\
\hline
HCOOH$^a$  & 6.7(-17)           & ---               & 2.8(-18)\\
HCOOH:H$_2$O 7:93\%$^b$ & ---          & 1.5(-17)          & ---\\
HCOOH--5H$_2$O$^c$ & 8.5(-17)       & 5.0(-17)      & 1.1(-18)\\
HCOOH--6H$_2$O$^c$ & 7.7(-17)       & 4.3(-17)      & 3.5(-18)\\
\hline
\end{tabular}

$^a$ Laboratory gas phase data by \citet{marechal1987}.

$^b$ Laboratory solid state data by \citet{hudson1999}.

$^c$ Theoretical calculations by \citet{park2006}.

\end{table*}

To determine the relative band strengths, the $\nu_{\rm S}$(C=O),
$\nu_{\rm B}$(OH/CH), and $\nu_{\rm S}$(C--O) features of HCOOH as
well as the $\nu_{\rm B}$(OH) and libration ($\nu$(lib)) bands of
H$_2$O are integrated over the frequency ranges given in
Table~\ref{int}. The integration boundaries for H$_2$O are identical
to those used recently by \citet{oberg2007}. Since the $\nu_{\rm
B}$(OH) band at 930~cm$^{-1}$ (10.8~$\mu$m) and the OCO bending mode
($\nu_{\rm B}$(OCO)) at 705~cm$^{-1}$ (14.2~$\mu$m) of HCOOH overlap
with the H$_2$O $\nu$(lib) band at 750~cm$^{-1}$(13.3~$\mu$m), these
two features are subtracted before integration.

\begin{table}
\caption{Frequency ranges used for integration bounds.}\label{int}
\begin{center}
\begin{tabular}{ll}
\hline
\hline
Vibrational mode & Wavenumber \\
& (cm$^{-1}$) \\
\hline
\multicolumn{2}{c}{HCOOH}\\
\hline
$\nu_{\rm S}$(C=O)               & 1800--1550  \\
$\nu_{\rm B}$(OH/CH)           & 1450--1370 \\
$\nu_{\rm S}$(C--O)              & 1300--1150 \\
\hline
\multicolumn{2}{c}{H$_2$O}\\
\hline
$\nu_{\rm S}$(OH)              & 3700--3000 \\
$\nu$(lib)                     & 1000--500 \\
\hline
\end{tabular}
\end{center}
\end{table}

The relative band strengths of the HCOOH and H$_2$O spectral features
are calculated with respect to pure HCOOH and H$_2$O via:

\begin{equation}
\frac{A{\rm (HCOOH:H_2O=Y:1)}}{A {\rm_0 (HCOOH)}} = Y \times
\frac{\int_{\rm band} I{\rm (HCOOH:H_2O=Y:1)}}{\int I{\rm_0 (HCOOH)}}.
\end{equation}

\noindent Here Y is the HCOOH ice fraction, $\int_{\rm band} I{\rm
  (HCOOH:H_2O=Y:1)}$ the integrated absorbance of the band in the
mixture, $A{\rm_0 (HCOOH)}$ the band strength for pure HCOOH ice at
15~K, and $\int I{\rm_0 (HCOOH)}$ the integrated absorbance for pure
HCOOH ice at 15~K. A similar formula is used for H$_2$O. The resulting
relative band strengths are studied as a function of temperature (see
Sect.~\ref{bandstrength}). An extensive discussion of the
uncertainties of the band strengths is given by \citet{oberg2007}. In
short, the largest fraction of the uncertainties arises from
inaccuracies on the mixing ratio of the ice and yields relative
uncertainties that are below 10\%. Additionally, baseline subtraction
and deposition time inaccuracies are 1 to 2\%. A conservative approach
yields an overall uncertainty of $\sim$12\% in the resulting
calculations for the relative band strengths. Furthermore, it is
important to note that the relative uncertainty between different
temperatures for a single experiment amounts to $\sim$2\% as only
baseline inaccuracies contribute. Observed temperature trends are thus
much more certain than concentration dependencies.

The $\nu_{\rm S}$(C=O), $\nu_{\rm B}$(OH/CH) and $\nu_{\rm S}$(C--O)
bands of HCOOH are fitted with Gaussian line profiles for the
astrophysically most relevant 1800--1100 cm$^{-1}$ range. To fit the
asymmetric profile of the $\nu_{\rm S}$(C=O) stretch around
1700~cm$^{-1}$ two components are needed. It turned out to be
difficult to disentangle the properties of the absorption bands of
HCOOH in some mixtures, in particular those containing CO and
CO$_2$. In such cases average peak positions are given. In general the
uncertainties of peak positions and FWHMs are $\pm$1~cm$^{-1}$ and
$\pm$2~cm$^{-1}$, except for the $\nu_{\rm B}$(OH/CH) feature in the
9:91\% HCOOH:H$_2$O mixture, where the values are $\pm$2~cm$^{-1}$ and
$\pm$4~cm$^{-1}$, respectively.

\section{Results}
\label{res}

The following section summarizes the results for the pure and mixed
HCOOH ices at different temperatures. All spectra are available at
http://www.strw.leidenuniv.nl/$\sim$lab/databases/.

\subsection{Pure HCOOH ices}
\label{hcooh_res}

\begin{table*}
\caption{Solid state spectral features for HCOOH, H$_2$O, CO, CO$_2$,
and CH$_3$OH at 15~K. The symbols $\nu_{\rm S}$ and $\nu_{\rm B}$
indicate a stretching or bending mode, respectively. The gas phase
spectroscopic label is indicated as well. The astrophysically most
relevant features of HCOOH are marked in bold face.}\label{spectra}
\begin{center}
\begin{tabular}{lllll}
\hline
\hline
Wavenumber & Wavelength & Mode & Label & Comment\\
(cm$^{-1}$) & ($\mu$m) &  & & \\
\hline
\multicolumn{4}{l}{HCOOH} \\
\hline
3115 & \phantom{1}3.21 & $\nu_{\rm S}$(OH) & $\nu_1$ & has several sub-maxima\\
2953 & \phantom{1}3.39 & $\nu_{\rm S}$(CH) & $\nu_2$& \\
2754 & \phantom{1}3.63 & $\nu_{\rm S}$(OH) & $\nu_1$ & has several sub-maxima\\
2582 & \phantom{1}3.87 & $\nu_{\rm S}$(OH) & $\nu_1$ & has several sub-maxima\\
{\bf 1714} & {\bf\phantom{1}5.83} & {\bf $\nu_{\rm S}$(C=O)} & {\bf $\nu_3$} & out-of-phase between adjacent chains\\
{\bf 1650} & {\bf \phantom{1}6.06} & {\bf $\nu_{\rm S}$(C=O)}  & {\bf $\nu_3$} & in-phase between adjacent chains\\
{\bf 1387} & {\bf \phantom{1}7.21} & {\bf $\nu_{\rm B}$(OH) \& {\bf $\nu_{\rm B}$(CH)} } & {\bf $\nu_5$} \& {\bf $\nu_4$}\\
{\bf 1211} & {\bf \phantom{1}8.26} & {\bf $\nu_{\rm S}$(C--O)} & {\bf $\nu_6$} & \\
1073 & \phantom{1}9.32 & $\nu_{\rm B}$(CH) & $\nu_8$ & out-of-plane\\
\phantom{1}930  & 10.75 & $\nu_{\rm B}$(OH) & $\nu_5$ & out-of-plane\\
\phantom{1}705  & 14.18 & $\nu_{\rm B}$(OCO) & $\nu_7$\\
\hline
\multicolumn{4}{l}{H$_2$O}\\
\hline
3280 & \phantom{1}3.05 & $\nu_{\rm S}$(OH) & $\nu_1,\nu_3$$^a$ & \\
1655 & \phantom{1}5.88 & $\nu_{\rm B}$(OH) & $\nu_2$\\
\phantom{1}750  & 13.33 & $\nu$(lib)        \\ 
\hline
\multicolumn{4}{l}{CO}\\
\hline
2140 & \phantom{1}4.67 & $\nu_{\rm S}$(CO) & $\nu_1$\\
\hline
\multicolumn{4}{l}{C$^{18}$O$_2$}\\
\hline
3671 & \phantom{1}2.72   &  overtone &  $\nu_1$+$\nu_3$\\
3513 & \phantom{1}2.85   &  overtone & 2$\nu_2$+$\nu_3$\\
2310 & \phantom{1}4.33 & $\nu_{\rm S}$(CO) & $\nu_3$\\
\phantom{1}645  & 15.50 & $\nu_{\rm B}$(OCO) & $\nu_2$\\
\hline
\multicolumn{4}{l}{CH$_3$OH}\\
\hline
3251 & \phantom{1}3.08 & $\nu_{\rm S}$(OH) & $\nu_1$\\
2951 & \phantom{1}3.39 & $\nu_{\rm S}$(CH) & $\nu_2,\nu_9$\\
2827 & \phantom{1}3.54 & $\nu_{\rm S}$(CH) & $\nu_3$\\
1460 & \phantom{1}6.85 & CH$_3$ deformation and $\nu_{\rm B}$(OH) & $\nu_4,\nu_{10},\nu_5, \nu_6$$^b$\\
1130 & \phantom{1}8.85 & CH$_3$ rock & $\nu_7, \nu_{11}$\\
1026 & \phantom{1}9.75 & $\nu_{\rm S}$(CO) & $\nu_8$\\
\phantom{1}694  & 14.41 & torsion & $\nu_{12}$\\ 
\hline
\end{tabular}
\end{center}

$^a$ $\nu_1$ refers to the symmetric stretching mode and
$\nu_3$ to the anti-symmetric stretching mode of H$_2$O.

$^b$ $\nu_4$ and $\nu_{10}$ are anti-symmetric deformations, $\nu_5$
is a symmetric deformation of CH$_3$OH and $\nu_6$ is the OH bending
mode for CH$_3$OH.

\end{table*}

\begin{figure}
\centering
\includegraphics[width=9cm]{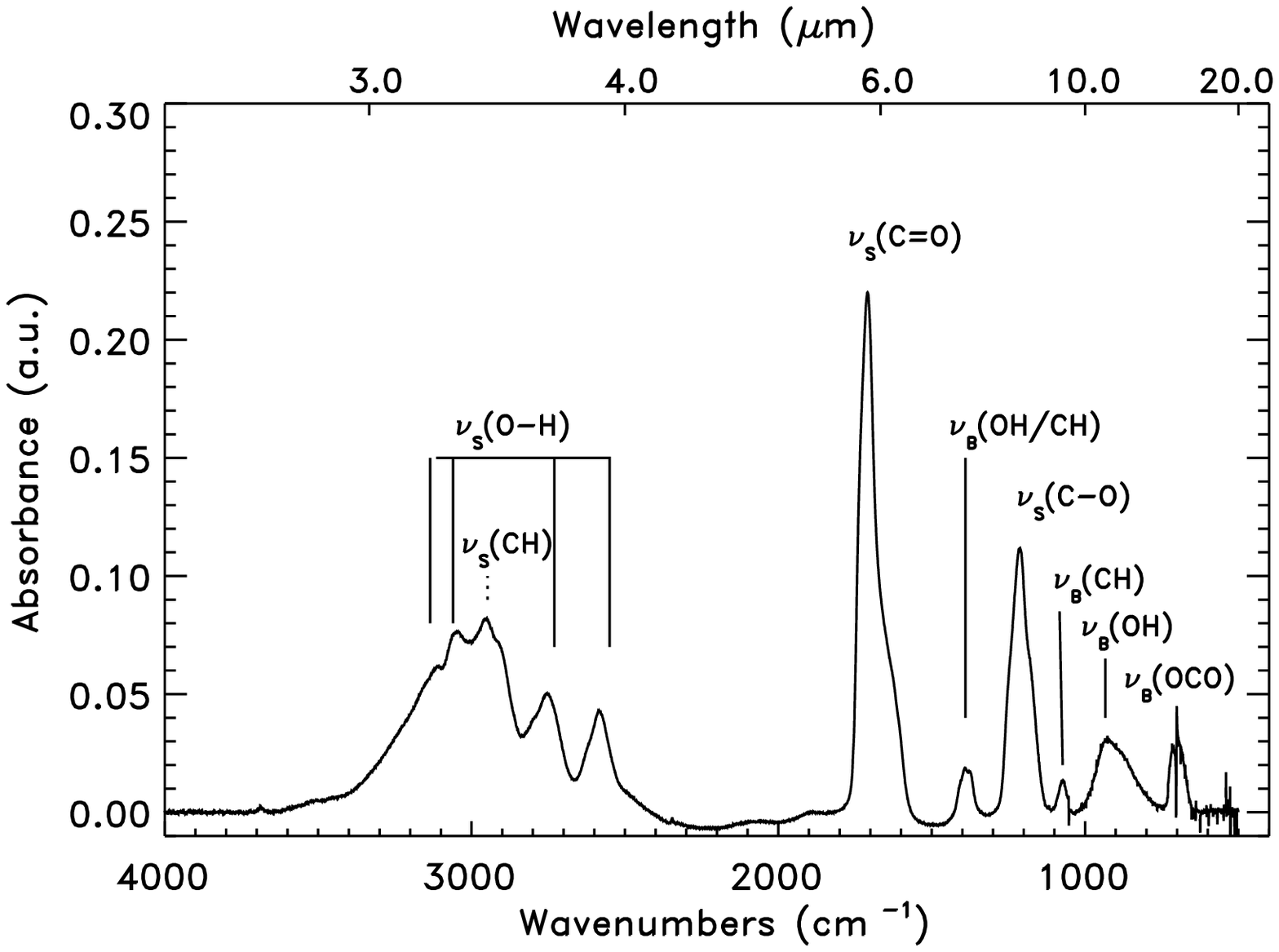}
\caption{Experimental infrared spectrum of pure HCOOH ice deposited at
15~K. The $\nu_{\rm S}$(OH) band has several sub-maxima in the
$\sim$3000~cm$^{-1}$ range where $\nu_{\rm S}$(CH) is also
located. Below 2000~cm$^{-1}$ the following absorption features are
present: the C=O stretching mode $\nu_{\rm S}$(C=O), the C--O stretch
$\nu_{\rm S}$(C--O), the CH bending mode $\nu_{\rm B}$(CH), the OH
bending mode $\nu_{\rm B}$(OH) and OCO bending mode $\nu_{\rm
B}$(OCO). Peak positions and labels are given in
Table~\ref{spectra}. In this study the focus is on the $\nu_{\rm
S}$(C=O), $\nu_{\rm B}$(OH/CH), and $\nu_{\rm S}$(C--O)
bands.}\label{spec1}
\end{figure}

Clearly, the pure HCOOH ice spectrum shown in Fig.~\ref{spec1} for
15~K is complex. For completeness we summarize the spectroscopic
assignments for all bands here (see Table \ref{spectra}), but in this
paper we mainly focus on the astrophysically relevant features
$\nu_{\rm S}$(C=O), $\nu_{\rm B}$(OH/CH), as well as $\nu_{\rm
  S}$(C--O). The measured peak positions and FWHMs in the
astrophysically relevant 1800-1100~cm$^{-1}$ range are given in
Table~\ref{pp_pure} of the online material. The features at
3115~cm$^{-1}$, 2754~cm$^{-1}$, and 2582~cm$^{-1}$ (3.21--3.87~$\mu$m)
are all due to the OH stretching mode, $\nu_{\rm S}$(OH). The broad
structure with different maxima of these modes are assigned to a
combination of dimers and HCOOH organized in long chains
\citep{cyriac2005}. The $\nu_{\rm S}$(OH) bands overlap with the
2953~cm$^{-1}$ CH stretch, $\nu_{\rm S}$(CH). In observations these
bands overlap with and are dominated by the 3~$\mu$m H$_2$O band. The
astrophysically important $\nu_{\rm S}$(C=O) band has its maximum
around 1714~cm$^{-1}$ and is broad and asymmetric. The C--O stretch,
$\nu_{\rm S}$(C--O), is located at 1211~cm$^{-1}$. Furthermore,
bending modes are present of which $\nu_{\rm B}$(CH) and $\nu_{\rm
  B}$(OH) at 1073~cm$^{-1}$ and 930~cm$^{-1}$ (9.3 and 10.8~$\mu$m),
respectively, are out-of-plane.

\begin{figure*}
\centering
\includegraphics[width=16cm]{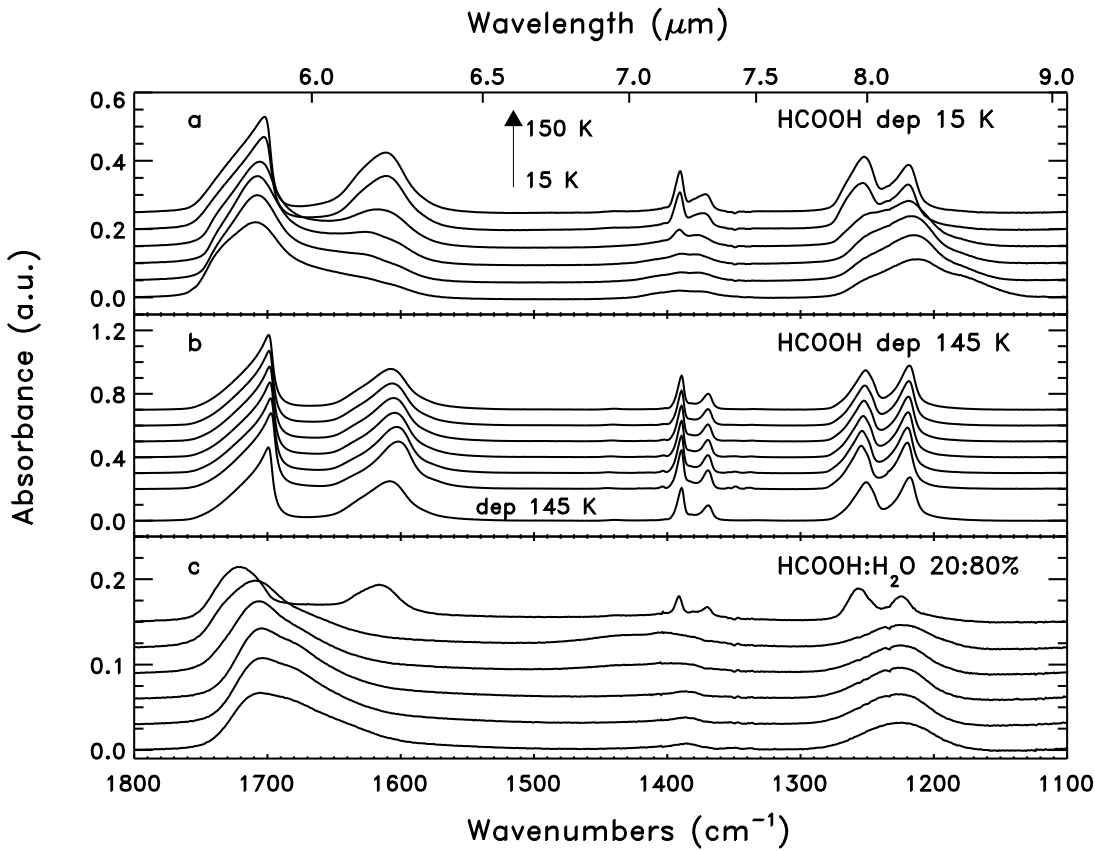}
\caption{The temperature dependent spectra of a) pure HCOOH ice
deposited at 15~K, b) pure HCOOH ice deposited at 145~K, subsequently
cooled down to 15~K, and c) HCOOH:H$_2$O 20:80\% mixed ice. The
spectra of the ices are shown for 15, 60, 90, 105, 135, and 150~K. The
$\nu_{\rm B}$(OH) band of H$_2$O at 1655~cm$^{-1}$ has been removed
from the spectra for the HCOOH:H$_2$O 20:80\%
mixture.}\label{purespec}
\end{figure*}

The temperature evolution of the astrophysically relevant HCOOH
features in the 1800--1100~cm$^{-1}$ range is shown in
Fig.~\ref{purespec}a and the corresponding peak positions are given in
Table~\ref{pp_pure} of the online material. Most features are broad
and consist of multiple components at 15~K. At 135~K the $\nu_{\rm
  S}$(C=O), $\nu_{\rm S}$(C--O), and $\nu_{\rm B}$(OH/CH) modes at
$\sim$1700, 1250, and 1390~cm$^{-1}$ split into two components. The
splitting of the $\nu_{\rm S}$(C=O) and $\nu_{\rm S}$(C--O) bands is
assigned to the out-of-phase and in-phase motions of different HCOOH
molecules within the same chain \citep{cyriac2005}. The out-of-phase
motions are located at higher wavenumbers than the in-phase
motions. The $\nu_{\rm S}$(OH) and $\nu_{\rm S}$(CH) in-plane bending
modes overlap at 15~K, but are clearly observed as two separate peaks
at higher temperatures. Our results are consistent with those of
\citet{cyriac2005}, who assigned the splitting to a conversion of
HCOOH dimers to HCOOH organized in chains, following and experimental
study by \citet{millikan1958} and calculations by
\citet{yokoyama1991}.

In one experiment HCOOH is deposited at 145~K (see
Fig.~\ref{purespec}b). After deposition the spectrum is similar to
that of HCOOH deposited at 15~K and subsequently heated to
$\geq$135~K. When the temperature is lowered to 15~K, the general
features remain similar to the high temperature spectrum. Upon
subsequent heating to 150~K an exact copy of the original (145~K)
spectrum is obtained. Thus, the changes in ice matrix structure after
heating are irreversible.

\subsection{HCOOH:H$_2$O ices}
\label{hcooh-h2o}

Experiments with mixtures that range from 50:50\% to 9:91\%
HCOOH:H$_2$O have been performed. In Figure~\ref{purespec}c, the
temperature evolution of the mixture with 20\% of HCOOH is shown. Peak
positions and FWHMs for the 1800--1100~cm$^{-1}$ range are given in
Table~\ref{pp_hcooh-h2o} of the online material. The dilution of HCOOH
in H$_2$O clearly affects the band profiles of HCOOH. The $\nu_{\rm
S}$(C=O) band shifts 3~cm$^{-1}$ to the red at 15~K and the $\nu_{\rm
B}$(OH/CH) feature is narrower than in pure HCOOH. Between 105 and
120~K both the peak intensity and FWHM of the $\nu_{\rm B}$(OH/CH)
band increase. At 150~K the H$_2$O features change due to a phase
transition from amorphous to crystalline H$_2$O ice structure. At the
same time the $\nu_{\rm S}$(C=O), $\nu_{\rm B}$(OH/CH), and $\nu_{\rm
S}$(C--O) modes of HCOOH split into two components. Finally, at 165~K
all H$_2$O and most HCOOH have desorbed.

\begin{figure}
\centering \includegraphics[width=9cm]{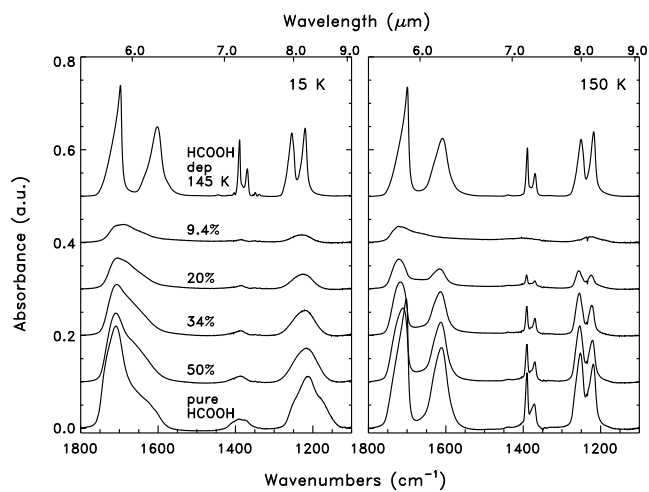}
\caption{Absorption spectra for temperatures of 15~K (left) and 150~K
(right) in the 1800--1100~cm$^{-1}$ range for HCOOH deposited at 15~K,
the HCOOH:H$_2$O mixtures and HCOOH deposited at 145~K. The $\nu_{\rm
B}$(OH) band of H$_2$O has been removed from the
spectra.\label{h2o_conc}}
\end{figure}

The spectra for different mixture concentrations with respect to
H$_2$O are shown in Fig.~\ref{h2o_conc} for 15~K (left) and 150~K
(right). The $\nu_{\rm S}$(C=O) band has an additional component at
15~K in HCOOH:H$_2$O mixtures with high HCOOH concentrations that is
located at lower wavenumbers and is due to the in-phase motion between
HCOOH molecules in the same chain. This implies that there is a small
fraction of HCOOH ice that is in a bulk HCOOH environment. The pure
HCOOH component decreases with increasing H$_2$O concentration. The
$\nu_{\rm B}$(OH/CH) and $\nu_{\rm S}$(C--O) modes consist of one
component and their FWHMs decrease for lower HCOOH concentrations.

At 150~K, the $\nu_{\rm S}$(C=O), $\nu_{\rm B}$(OH/CH), and $\nu_{\rm
S}$(C--O) modes are split into two peaks for the three mixtures with
the highest HCOOH concentrations. The $\nu_{\rm B}$(OH/CH) mode is
broader compared to pure HCOOH. At the lowest HCOOH concentration, no
splitting of the $\nu_{\rm S}$(C=O) spectral feature of HCOOH is
observed, but only a broad band with a FWHM of
67~cm$^{-1}$. Furthermore, the $\nu_{\rm S}$(C=O) band is shifted to
1718~cm$^{-1}$ and the $\nu_{\rm B}$(OH/CH) feature is located at
1404~cm$^{-1}$.

\subsection{HCOOH:CO, HCOOH:CO$_2$, and HCOOH:CH$_3$OH ices}
\label{mix_sec}
\begin{figure*}
\centering
\includegraphics[width=16cm]{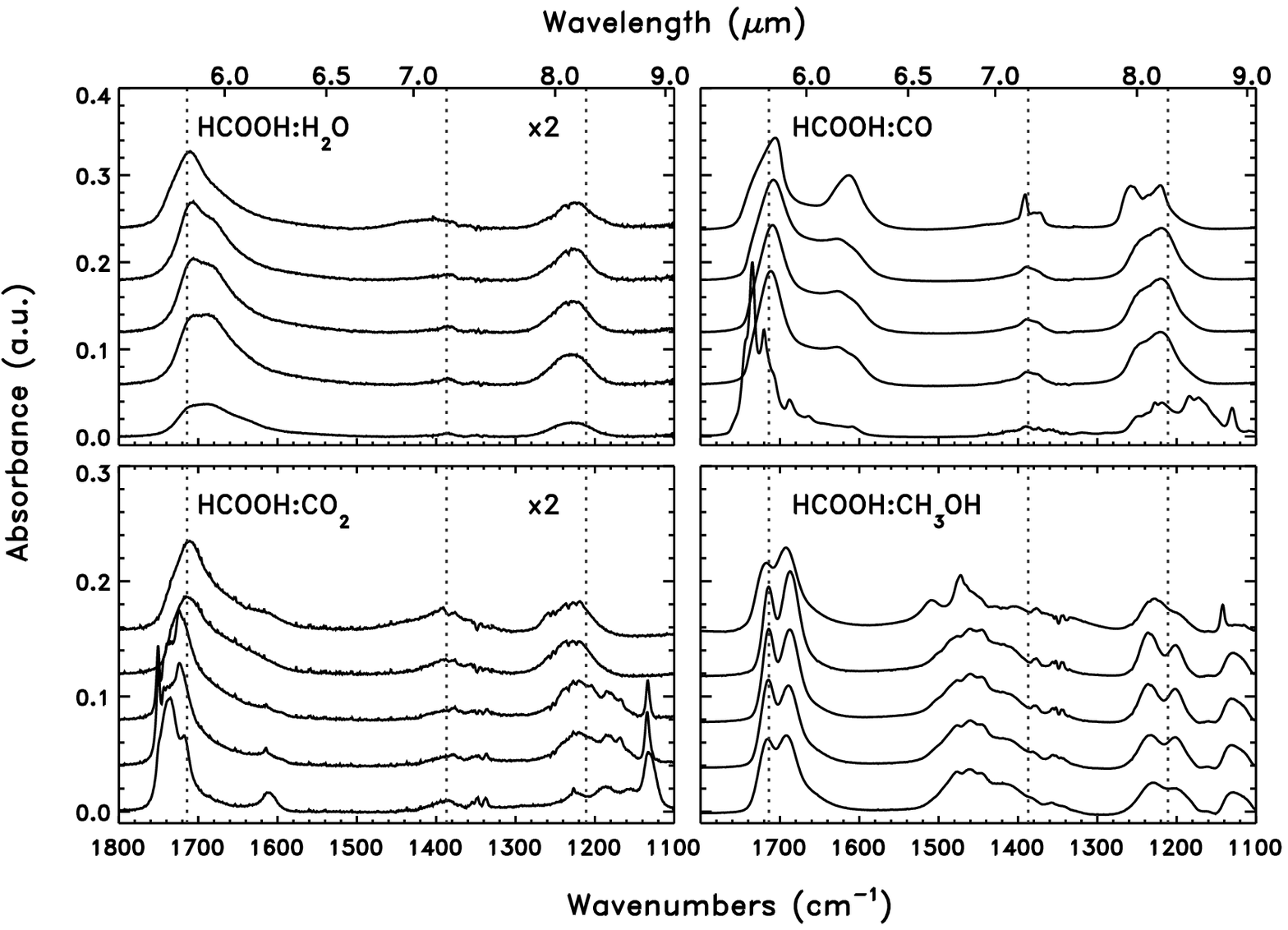}
\caption{Absorption spectra in the 1800-1100~cm$^{-1}$ regime for
mixtures with $\sim$90\% H$_2$O, CO, CO$_2$ and CH$_3$OH at 15 (bottom
curve), 45, 75, 105, and 135~K (top curve). The $\nu_{\rm B}$(OH) has
been removed for the HCOOH:H$_2$O mixture. The peak positions of pure
HCOOH at 15~K are indicated by the dotted lines. The intensities of
the bands are multiplied by a factor 2 for the HCOOH:H$_2$O and
HCOOH:CO$_2$ mixtures as indicated in the first and third panel,
respectively.\label{solvT}}
\end{figure*}

Figure~\ref{solvT} shows the effects different species have on the
solid state HCOOH infrared spectrum. Tables~\ref{pp_hcooh-ch3oh} and
\ref{nuc=o} of the online material summarize the profile
parameters. HCOOH mixed with H$_2$O results in a relatively simple
spectrum, but mixtures with other species, especially CO or CO$_2$,
show multiple components. In mixtures with CO or CO$_2$ the $\nu_{\rm
S}$(C=O) band is shifted to higher wavenumbers of which the two most
prominent features are located at 1717~cm$^{-1}$ and 1735~cm$^{-1}$ in
both mixtures. A feature is present at 1154~cm$^{-1}$ for the
$\nu_{\rm S}$(C--O) band which is assigned to monomeric HCOOH. This
band is especially strong in the HCOOH:CO$_2$ mixture. In CH$_3$OH,
the absorption bands more closely resemble the overall features of
mixtures with H$_2$O. However, the $\nu_{\rm S}$(C=O) and $\nu_{\rm
S}$(C--O) modes are split into two components at 1721~cm$^{-1}$ and
1691~cm$^{-1}$. The $\nu_{\rm B}$(OH/CH) band cannot be compared to
that in the other mixtures as it overlaps with and is dominated by the
CH$_3$ deformation and $\nu_{\rm B}$(OH) features of CH$_3$OH at
1460~cm$^{-1}$.

The temperature behavior of the spectral features of HCOOH ice depends
strongly on the matrix species. In mixtures with H$_2$O, the HCOOH
bands are similar for all temperatures. This is not surprising as the
H$_2$O ice structure does not change until 135~K. HCOOH behaves
differently in mixtures with CO and CO$_2$ at high temperatures,
however. When CO has desorbed at 45~K, the $\nu_{\rm S}$(C=O) and
$\nu_{\rm S}$(C--O) features are similar to pure HCOOH. Furthermore,
its integrated intensity does not decrease after CO desorption. In the
mixture of HCOOH with CO$_2$, the 1735~cm$^{-1}$ and 1717~cm$^{-1}$
bands of HCOOH only fully disappear between 75~K and 105~K at the same
time that the features for pure HCOOH start to appear and after CO$_2$
has desorbed. Thus even when HCOOH is present as a minor component in
CO or CO$_2$ it does not co-desorb at low temperatures. The HCOOH
bands in CH$_3$OH have a similar lack of temperature variation as for
mixtures with H$_2$O. Between 105 and 135~K CH$_3$OH has a phase
transition that is observed around 1500~cm$^{-1}$ where both the
CH$_3$ deformation and $\nu_{\rm B}$(OH) modes of CH$_3$OH are
located. The HCOOH $\nu_{\rm S}$(C=O) and $\nu_{\rm S}$(C--O) bands do
not resemble pure HCOOH features at 135~K, which indicates that both
species are still mixed.

\subsection{Tertiary mixtures}
\label{tert}

Spectra of tertiary mixtures $\sim$7:67:26\% of HCOOH:H$_2$O:CO,
HCOOH:H$_2$O:CO$_2$ and HCOOH:H$_2$O:CH$_3$OH have been measured at
15~K. In Table~\ref{ppmix} and Table~\ref{nuc=o} of the online
material peak positions and FWHMs are given. The $\nu_{\rm S}$(C=O)
and $\nu_{\rm S}$(C--O) features are similar to those in mixtures with
H$_2$O, but are shifted by a few wavenumbers. The $\nu_{\rm B}$(OH/CH)
band is located at lower wavenumbers, especially for the
HCOOH:H$_2$O:CH$_3$OH mixture and have FWHMs of
$\sim$14--19~cm$^{-1}$.

\begin{table}
\caption{The peak positions, FWHMs, and relative band strengths of
$\nu_{\rm B}$(OH/CH) at 15~K of tertiary mixtures of HCOOH:H$_2$O:X,
where X stands for CO, CO$_2$, and CH$_3$OH. The uncertainties on the
peak positions and FWHMs are $\pm$1~cm$^{-1}$ and $\pm$2~cm$^{-1}$,
respectively.}\label{ppmix}
\begin{tabular}{llll}
\hline
\hline
Mixture & Peak position & FWHM & $A/A_{\rm 0}$\\
        & (cm$^{-1}$)   & (cm$^{-1}$) & \\
\hline
HCOOH:H$_2$O:CO & 1386 & 19   & 4.2\\
HCOOH:H$_2$O:CO$_2$ & 1386 & 19  & 3.0\\
HCOOH:H$_2$O:CH$_3$OH & 1382 & 13 & 8.3\\ 
\hline
\end{tabular}
\end{table}

\subsection{Comparison to previous laboratory data}
\label{prev-lab}

Prior to this study, only \citet{schutte1999} and \citet{hudson1999}
reported infrared spectra of HCOOH mixtures with H$_2$O and CH$_3$OH
at 1 and 4~cm$^{-1}$ resolution, respectively. Their experiments only
focused on the $\nu_{\rm B}$(OH/CH) and $\nu_{\rm S}$(C--O) bands and
consequently it is difficult to make a full comparison with our data
which also includes other HCOOH bands. The peak positions of the
$\nu_{\rm B}$(OH/CH) features in pure HCOOH ice and 10:100\%
HCOOH:H$_2$O mixtures by \citet{schutte1999} are shifted and the FWHMs
are narrower compared to our data. However, the tertiary
HCOOH:H$_2$O:CH$_3$OH 8:66:26\% mixture results agree very well within
the experimental uncertainties. These discrepancies are likely due to
a different baseline subtraction. In pure HCOOH and mixtures with
H$_2$O the $\nu_{\rm B}$(OH/CH) mode has a wing on the blue side,
which is not subtracted in the present work. The peak position and
FWHM for the $\nu_{\rm S}$(C--O) feature for a 7:93\% HCOOH:H$_2$O
mixture by \citet{hudson1999} are the same within the uncertainties.

\section{Band strength changes in ice mixtures}
\label{bandstrength}

As stated before in Sect.~\ref{analysis} absolute band strengths are
difficult to measure in our experiment, whereas the relative band
strengths can be accurately determined. The relative values discussed
in this section can be used in combination with existing values in the
literature given in Table~\ref{ab_band} to derive absolute band
strengths in different ice environments.

\subsection{HCOOH}

\begin{figure}
\centering
\includegraphics[width=9cm]{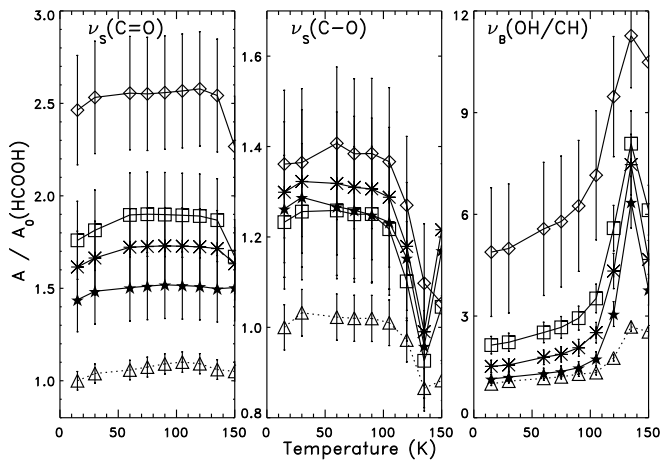}
\caption{$A/A_0{\rm ( HCOOH)}$ with respect to temperature for the
  HCOOH $\nu_{\rm S}$(C=O), $\nu_{\rm B}$(OH/CH), and $\nu_{\rm
    S}$(C--O) bands in HCOOH:H$_2$O. The $\star$ symbols refer to the
  50:50\% mixture, $\ast$ to 34:66\%, $\Box$ to 20:80\%, $\Diamond$ to
  9:91\% HCOOH:H$_2$O, and $\bigtriangleup$ to pure HCOOH. Note the
  different scales on the vertical axes.}
\label{int_comp}
\end{figure}

When HCOOH ice is mixed with H$_2$O the band strengths of HCOOH are
strongly affected. In Figure~\ref{int_comp} the relative band
strengths of the $\nu_{\rm S}$(C=O), $\nu_{\rm S}$(C--O) and $\nu_{\rm
  B}$(OH/CH) spectral features are shown with error-bars of 12\%
superimposed. Clearly the uncertainties are relatively large, but
still it is possible to note a number of effects with concentration
and temperature. As discussed in Sect.~\ref{analysis} the error-bars
in the same experiment at different temperatures are only 2\%, and
observed temperature trends are thus much more accurate than
concentration trends.

All mixtures with H$_2$O show a similar temperature trend as pure
HCOOH. Even though between 135 and 150~K the $\nu_{\rm S}$(C=O) band
splits, its band strength is constant with temperature. In contrast,
that of the $\nu_{\rm S}$(C--O) feature shows a pronounced dip around
135~K, which is partly recovered at 150~K. For the most diluted
mixture where chain-formation does not occur, no such dip is
present. The band strength of the $\nu_{\rm B}$(OH/CH) mode increases
with temperature and peaks at 135~K. A striking difference between the
three absorption bands is that the band strength increases by a factor
2 to 5 with decreasing HCOOH ice concentration for the $\nu_{\rm
S}$(C=O) and $\nu_{\rm B}$(OH/CH) bands, but is the same within 30\%
for the $\nu_{\rm S}$(C--O) feature in all mixtures.

\begin{figure}
\centering
\includegraphics[width=9cm]{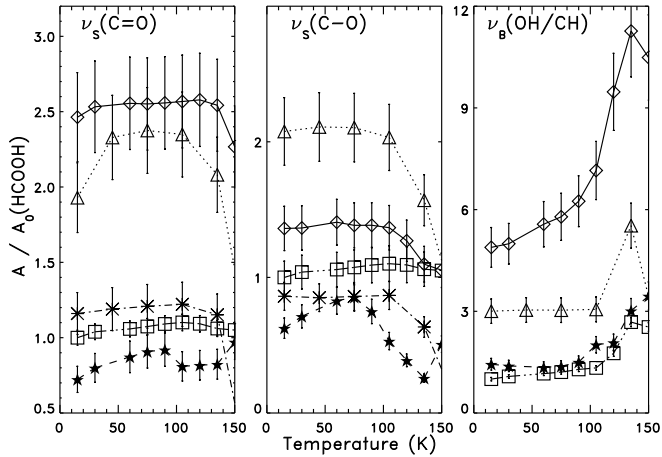}
\caption{$A/A_0{\rm (HCOOH)}$ with respect to temperature for the
HCOOH $\nu_{\rm S}$(C=O), $\nu_{\rm B}$(OH/CH), and $\nu_{\rm
S}$(C--O) bands. The $\Box$ symbols refers to pure HCOOH, $\Diamond$
to 9:91\% HCOOH:H$_2$O, $\bigtriangleup$ to 11:89\% HCOOH:CO, $\star$
to 9:91\% HCOOH:CO$_2$, and $\ast$ to 10:90\% HCOOH:CH$_3$OH. Note the
different scales on the vertical axes.}
\label{int_comp_mat}
\end{figure}

A similar comparison for the relative band strength with respect to
temperature is made for mixtures with CO, CO$_2$, and CH$_3$OH (see
Fig.~\ref{int_comp_mat}). The band strengths in mixtures with CO are
higher compared to those in pure HCOOH, even when all CO has
evaporated. However, the band strengths in CO$_2$ are close to or
smaller than those for pure HCOOH. The $\nu_{\rm S}$(C=O) and
$\nu_{\rm S}$(C--O) modes are largest around 75--90~K, decrease
between 90--135~K and finally increase again up to 150~K. The
$\nu_{\rm S}$(C=O) and $\nu_{\rm S}$(C--O) bands are relatively little
affected in the mixtures with CH$_3$OH.

The relative band strengths for the $\nu_{\rm B}$(OH/CH) mode in the
tertiary mixtures are shown in Table~\ref{ppmix} and those of the
other bands in Table~\ref{nuc=o} of the online material. Clearly, the
band strength in the tertiary mixtures with H$_2$O and CO, CO$_2$ or
CH$_3$OH are enhanced at 15~K.

In summary, the relative band strengths of the $\nu_{\rm S}$(C=O) and
$\nu_{\rm S}$(C--O) bands can range from 0.8--3~$A_0$(HCOOH) and of
the $\nu_{\rm B}$(OH/CH) band from 1--12~$A_0$(HCOOH) in different ice
environments.

\subsection{H$_2$O}

\begin{figure}
\centering
\includegraphics[width=9cm]{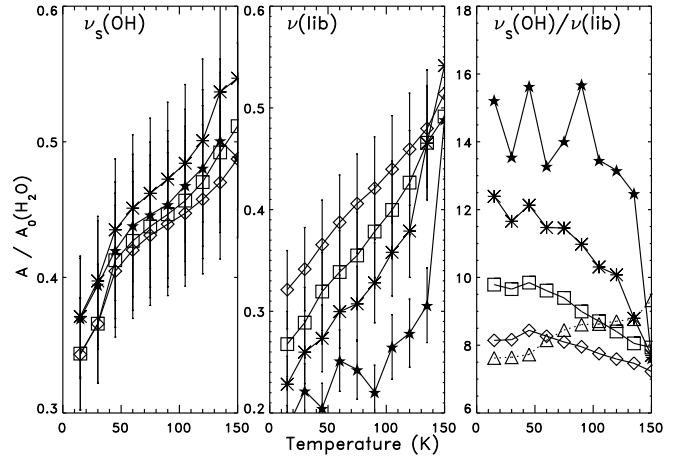}
\caption{$A/A_0{\rm ( H_2O)}$ of the H$_2$O $\nu_{\rm S}$(OH),
  $\nu$(lib), and the band strength ratio $A_{\nu\rm_S (OH)}/A_{\rm
    lib}$ with respect to temperature in HCOOH:H$_2$O mixtures and
  pure H$_2$O. The dotted line with $\bigtriangleup$ in the $A_{\rm
    \nu_S(OH)}/A_{\rm lib}$ panel indicates the trend for pure H$_2$O,
  the $\star$ for the 50:50\% mixture, $\ast$ for 34:66\%, $\Box$ for
  20:80\%, and $\Diamond$ for 9:91\% HCOOH:H$_2$O. Note the different
  scales on the vertical axes.}\label{int_comp_h2o}
\end{figure}

Water absorption bands are strongly affected by the presence of other
ice components like CO$_2$ \citep{oberg2007}. In
Figure~\ref{int_comp_h2o} the relative band strengths compared to pure
H$_2$O of the $\nu_{\rm S}$(OH) and the $\nu$(lib) modes in the
mixtures with HCOOH are shown versus temperature. The $\nu_{\rm
  S}$(OH) band strength is significantly smaller in all mixtures
compared to that of pure H$_2$O, but does not depend on the mixing
ratio. The $\nu$(lib) mode behaves differently. It decreases as a
function of HCOOH concentration, with the smallest band strength for
the highest HCOOH concentrations. As the temperature increases the
relative band strength of the $\nu$(lib) band increases, until a
temperature of 135~K is reached. The different behavior of $\nu_{\rm
  S}$(OH) and the $\nu$(lib) mode are further illustrated in the right
panel of Fig.~\ref{int_comp_h2o}. For pure H$_2$O the absorbance ratio
$A_{\nu\rm_S (OH)}/A_{\nu\rm (lib)}$ (the 3.0/13~$\mu$m features)
starts at $\sim$7.5 and increases with temperature to 9.5. The
mixtures have a higher ratio at 15~K, but decrease with
temperature. The $A_{\nu\rm_S (OH)}/A_{\nu\rm (lib)}$ ratio
furthermore increases with HCOOH concentration at 15~K. In conclusion,
the ratio of the relative band strengths of the 3.0/13~$\mu$m features
with a value larger than 10 implies that H$_2$O is present in ice with
HCOOH concentrations $>$20\%, and a lower value implies that the HCOOH
concentration is $<$20\%.

\section{Discussion}
\label{chemdisc}

In this section, our results are discussed in the light of recent
chemical physics experiments of HCOOH-containing ices. The behavior of
HCOOH in different ices may lead to important clues regarding the
behavior of the species under interstellar conditions. Of particular
relevance for interstellar chemistry is the mobility in the ice and
desorption behavior of the ice constituents with temperature.

The infrared spectrum of pure HCOOH distinguishes itself from HCOOH
mixtures by the presence of dimers to a chain-like structure of HCOOH
molecules at 120--135~K for HCOOH deposited on a KBr substrate at 15~K
\citep[see e.g.,][]{cyriac2005}. The chain-structure at higher
temperatures results in a spectrum in which the $\nu_{\rm S}$(C=O) and
$\nu_{\rm S}$(C--O) bands are split. The transition from dimers to
HCOOH-chains is irreversible as evidenced by the splitting of the
$\nu_{\rm S}$(C=O) mode at 15~K for HCOOH deposited at 145~K (see
Fig.~\ref{purespec} and Sect.~\ref{hcooh_res}).

In binary mixtures with H$_2$O, the HCOOH ice spectral features remain
relatively similar throughout a large temperature regime. Only from
120~K onwards, changes are observed for our experiments (see
Fig.~\ref{h2o_conc}). Initially, only the $\nu_{\rm B}$(OH/CH) mode
increases in intensity and broadens, but between 135~K and 150~K the
$\nu_{\rm S}$(C=O) band splits into the out-of-phase and in-phase
modes for $>$10\% HCOOH, similar to what is observed for pure HCOOH
ice. This implies that the molecules first reorganize locally at 120~K
and subsequently become mobile throughout the ice at higher
temperatures. This is corroborated by studies of layered binary
HCOOH-H$_2$O ices
\citep{cyriac2005,borodin2005,bahr2005,souda2006}. In addition to the
infrared experiments by \citet{cyriac2005}, Metastable Impact Electron
Spectroscopy (MIES) and Temperature Programmed Desorption (TPD) have
been used to probe the layered HCOOH/H$_2$O ices. No mixing is
observed in any of these experiments up to 120~K. Furthermore, the
splitting of the $\nu_{\rm S}$(C=O) mode between 135~K and 150~K in
our experiments implies that H$_2$O and HCOOH segregate at high
temperatures. This is consistent with TPD data from \citet{bahr2005},
where HCOOH desorbs after H$_2$O, even in ices where H$_2$O has been
deposited on top of HCOOH.

The interactions between HCOOH and CO, CO$_2$ and CH$_3$OH are
different compared to H$_2$O. Very little has been published about
either dimers or ice mixtures of these species. \citet{park2002}
performed experiments with HCOOH dissolved in liquid CO$_2$ and
C$_2$H$_6$. The dimer/monomer ratio for HCOOH was higher in C$_2$H$_6$
than CO$_2$. This is due to the stronger electrostatic interactions of
HCOOH with CO$_2$ compared to C$_2$H$_6$.  A similar effect may be
responsible for the difference between CO and CO$_2$ as the monomeric
HCOOH feature at 1154~cm$^{-1}$ is stronger in the mixture with CO$_2$
compared to CO. When HCOOH is diluted in an ice dominated by other
molecules with no or very small dipole-moments like CH$_4$, N$_2$ and
O$_2$, a similar HCOOH ice spectrum is likely obtained. The
monomer/dimer ratio will depend on the HCOOH concentration as well as
the strength of the electrostatic interaction between HCOOH and the
other ice species. A similar behavior is observed for H$_2$O ice
diluted in CO$_2$, CO, N$_2$ and O$_2$ (e.g., \citealp{oberg2007};
Bouwman et al., in prep.; Awad et al., in prep.).

The CH$_3$OH spectrum is very little affected by HCOOH. This is
probably due to a comparable strength of hydrogen bonds between HCOOH
and CH$_3$OH. The infrared spectra of HCOOH:CH$_3$OH 10:90\% mixtures
shown in Fig.~\ref{solvT} clearly illustrate that both species do not
segregate at temperatures below 135~K, even when CH$_3$OH has a phase
transition. It is likely that HCOOH and CH$_3$OH only become mobile
between 135~K and 150~K, as is found for mixtures of HCOOH with
H$_2$O. Since CH$_3$OH has a smaller dipole-moment and fewer
possibilities for making hydrogen-bonds compared to H$_2$O, it is also
expected to segregate from, and desorb prior to, HCOOH. Other polar or
hydrogen-bonding species such as NH$_3$ and H$_2$CO may interact with
HCOOH and affect its infrared spectrum in the same way.

\section{Astrophysical implications}
\label{astro}

\begin{figure*}
\includegraphics[width=18cm]{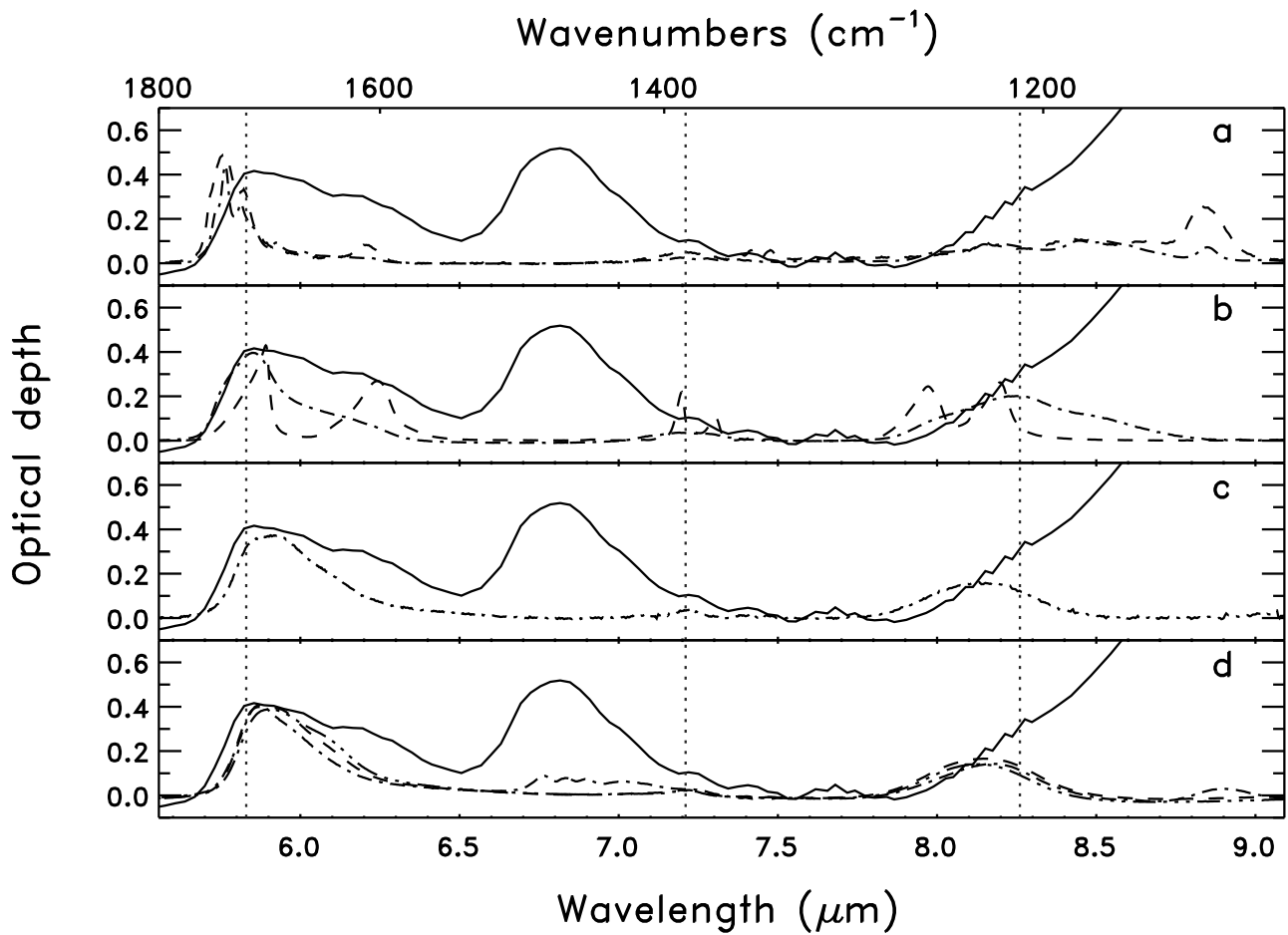}
\caption{The observed ice spectrum toward HH~46 in the 5--9~$\mu$m
  range \citep{boogert2004}. Also included are in a) are the
  $\sim$10:90\% mixtures of HCOOH:CO (dot-dash) and HCOOH:CO$_2$
  (dashed), in b) pure HCOOH spectra at 15~K for ice deposited at 15~K
  (dot-dash) and at 145~K (dashed), in c) the $\sim$10:90\% mixtures
  of HCOOH:H$_2$O (dash-dot) and CH$_3$OH (dash-dot-dot-dot) and in d)
  the tertiary mixtures 7:67:26\% of HCOOH:H$_2$O:CO
  (dash-dot-dot-dot), HCOOH:H$_2$O:CO$_2$(dashed), and
  HCOOH:H$_2$O:CH$_3$OH (dot-dash) at 15~K. The relevant peak
  positions of pure HCOOH ice at 15~K are indicated with dotted lines
  to guide the eye.}\label{obsa}
\end{figure*}

\subsection{HCOOH ice identification}
\label{id}

Formic acid ice has been observed in many different astrophysical
objects, such as high mass and low mass star-forming regions (e.g.,
\citealp{schutte1997,schutte1999}; Boogert et al., in prep.). In this
section we compare the laboratory spectra with the observed ISO
spectrum toward the low mass YSO HH~46 and the high mass YSO W~33A
\citep{boogert2004,gibb2000b}. The 5.9~$\mu$m ($\nu_{\rm S}$(C=O))
7.2~$\mu$m ($\nu_{\rm B}$(OH/CH)) and 8.1~$\mu$m ($\nu_{\rm S}$(C--O))
bands are used to determine the the ice environment in which HCOOH is
located. The 5.9~$\mu$m HCOOH feature is blended with other
interstellar bands and the 8.1~$\mu$m band overlaps with the wing of a
strong silicate feature. Nonetheless they can be used to exclude some
astrophysical HCOOH ice environments. In Figure~\ref{obsa} the
5--9~$\mu$m range is shown toward the low mass YSO HH~46 with the
experimental spectra over-plotted. In Figure~\ref{obsa}a the spectra
for the $\sim$10:90\% HCOOH:CO and HCOOH:CO$_2$ mixtures are
shown. These mixtures can clearly be ruled out for HH~46 because the
$\nu_{\rm S}$(C=O) feature is shifted too far to the blue. Similarly,
pure high temperature HCOOH (Fig.~\ref{obsa}b) can be discarded as the
out-of-phase $\nu_{\rm S}$(C--O) mode at 8.1~$\mu$m band is not
detected. Furthermore, for all pure ices and mixtures except the
binary HCOOH:CH$_3$OH ice, the $\nu_{\rm S}$(C--O) band causes a small
wing on the silicate band at 8.1~$\mu$m, which is not observed due to
the much larger intensity of the silicate absorption.

\begin{figure}
\centering
\includegraphics[width=9cm]{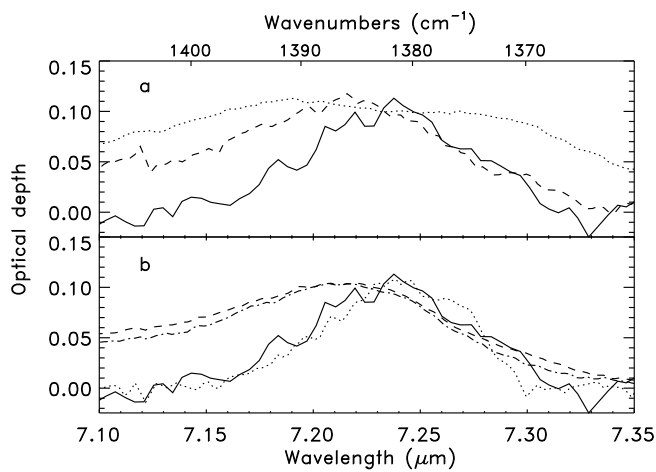}
\caption{The observed ice spectrum in the 7.10--7.35~$\mu$m range
  toward W~33A indicated by the solid line \citep{gibb2000b}. Also
  included are in a) the spectra at 15~K of pure HCOOH (dotted) and
  20:80\% HCOOH:H$_2$O (dashed) and in b) the tertiary mixtures
  HCOOH:H$_2$O:CO 8:62:30\% (dashed), HCOOH:H$_2$O:CO$_2$ 6:67:27\%
  (dash-dot) and HCOOH:H$_2$O:CH$_3$OH 6:68:26\% (dotted).}\label{obs}
\end{figure}

The integrated absorbance and peak position is determined most
accurately from the $\nu_{\rm B}$(OH/CH) band, because it is the least
blended feature of HCOOH. In Figure~\ref{obs} the spectrum observed
with ISO toward W~33A is shown for the $\nu_{\rm B}$(OH/CH)
feature. The ISO data are chosen for comparison because they have
higher spectral resolution ($\lambda / \Delta \lambda \approx$~800, or
2~cm$^{-1}$ at 7.2~$\mu$m) than the Spitzer data ($\lambda / \Delta
\lambda \approx$~100, or 14~cm$^{-1}$ at 7.2~$\mu$m). In
Figure~\ref{obs}a the 15~K spectra of pure HCOOH and 20:80\%
HCOOH:H$_2$O are shown. Clearly, the pure HCOOH ice spectrum is too
broad to fit the astronomical feature. The $\nu_{\rm B}$(OH/CH) mode
is also too far shifted to the blue for the HCOOH:H$_2$O mixture. The
tertiary mixtures depicted in Fig.~\ref{obs}b match the observations
better. The HCOOH spectral features in the HCOOH:H$_2$O:CO and
HCOOH:H$_2$O:CO$_2$ mixtures are almost identical, but have a blue
wing that is not observed for W~33A. However, the
HCOOH:H$_2$O:CH$_3$OH 6:68:26\% mixture reproduces the observed
spectrum very well.

\begin{figure}
\centering
\includegraphics[width=8cm]{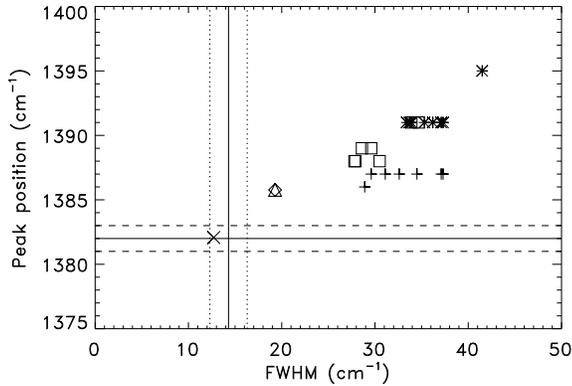}
\caption{Comparison between the observed 7.2~$\mu$m band of HCOOH in
  W~33A with laboratory spectra of HCOOH ice. The black lines indicate
  the peak position and FWHM of the observed feature and the dotted
  and dashed lines 2~cm$^{-1}$ uncertainties on the FWHM and
  1~cm$^{-1}$ on the peak position, respectively. Mixture ratios for
  all temperatures have the same symbol. The $+$ symbols refer to pure
  HCOOH, $\ast$ to 50:50\% HCOOH:H$_2$O, $\Box$ to 9:91\%
  HCOOH:H$_2$O, $\bigtriangleup$ to HCOOH:H$_2$O:CO 8:62:30\%,
  $\Diamond$ to 6:67:27\% HCOOH:H$_2$O:CO$_2$ and $\times$ to
  6.5:26:68\% HCOOH:CH$_3$OH:H$_2$O.}\label{peakshape}
\end{figure}

This sensitivity of peak position and FWHM of the $\nu_{\rm B}$(OH/CH)
band at 7.2~$\mu$m is further illustrated in Fig.~\ref{peakshape},
together with the observed values toward the high mass YSO W~33A. A
Gaussian fit is made to the observations in the same way as to the
experimental data. The black horizontal line corresponds to the fitted
peak position in the observations and the black vertical line to the
fitted FWHM. The dotted and dashed lines represent 2~cm$^{-1}$ and
1~cm$^{-1}$ error-bars for the FWHM and peak position,
respectively. Clearly, the binary mixtures with H$_2$O as well as pure
HCOOH are too broad and shifted too far to the blue side compared to
the observed feature in W~33A. The tertiary mixtures provide better
fits. The 6:27:67\% HCOOH:CO$_2$:H$_2$O and 8:62:30\% HCOOH:H$_2$O:CO
mixtures have FWHMs that are too broad compared to the observations
within their uncertainties. Furthermore, their peak positions are
shifted to 1386~cm$^{-1}$, a few wavenumbers too high. Potentially,
CO$_2$/CO:H$_2$O ratios higher than 0.40-0.52:1 could shift the
$\nu_{\rm B}$(OH/CH) feature to lower wavenumbers and match the
observations toward W~33A better. This is not unlikely as
\citet{oberg2007} find that they can fit the observed Spitzer spectrum
toward HH~46 well if also a 0.5:1 CO$_2$:H$_2$O mixture is
included. The 7.2~$\mu$m band for the HCOOH:CH$_3$OH:H$_2$O mixture
fits the observed spectrum very well.

\subsection{HCOOH ice formation}

The presence of HCOOH in a complex mixture dominated by H$_2$O but
also including CO, CO$_2$ or CH$_3$OH or a combination of these three
species is not unlikely when potential formation mechanism are
evaluated. CO or CO$_2$ are both potential precursors to HCOOH and
CH$_3$OH is known to form from CO through H-atom bombardment
(\citealp{watanabe2004}; Fuchs et al., in prep.). H-bombardment of CO
or CO$_2$ in mixtures with H$_2$O (Fuchs et al., in prep.; Bisschop et
al., in prep.) and UV-irradiation of CO:H$_2$O and CO$_2$:H$_2$O
mixtures \citep{watanabe2004,watanabe2002,gerakines2000} do not give
HCOOH as one of the main products. Since the UV experiments were
performed under HV conditions, they need to be repeated under ultra
high vacuum conditions to fully exclude this route. In H$_2$O:CO
mixtures that are bombarded by 0.8~MeV protons, to simulate cosmic
rays, HCOOH is formed \citep{hudson1999}. Its formation is explained
by:

\noindent H $+$ CO $\rightarrow$  HCO

\noindent OH $+$ HCO $\rightarrow$ HCOOH

\noindent where the H and OH result from H$_2$O dissociation in the
experiment. In space, H-atoms from the gas phase can react with the
adsorbed O to form OH. HCO is formed by H addition to CO, which will
further react to H$_2$CO and finally to CH$_3$OH, unless HCO
encounters an OH radical first. \citet{garrod2006} use this formation
mechanism for HCOOH in the solid state. However, in their model HCOOH
can only form at higher temperatures because the OH and HCO radicals
are not mobile at low temperatures. This would be inconsistent with
the detection of HCOOH ice in cold sources such as HH~46 and
background stars unless OH is produced with excess energy. This is for
example found in theoretical calculations by \citet{andersson2006} for
the photo-dissociation of H$_2$O ice. The newly formed OH radicals are
not very mobile in bulk ice, but can move more than 80~\AA\ over the
surface. Thus HCOOH formation through this reaction mechanism is
expected to be strongly localized at the ice surface. Another
potential formation mechanism is given by \citet{keane2001}.

\noindent HCO $+$ O $\rightarrow$ HCOO

\noindent HCOO $+$ H $\rightarrow$ HCOOH

\noindent In this case successive H- and O-addition leads to HCOOH
formation, and HCOOH may be produced at low ice temperatures at the
same time as H$_2$O and CH$_3$OH are formed.

In summary, both reaction pathways make it likely that HCOOH is formed
in an environment where H$_2$O and CH$_3$OH are located. Potentially
CO, CO$_2$, and H$_2$CO are present in the ice as well, if conversion
to CH$_3$OH and HCOOH is incomplete. As discussed in Sect.~\ref{id} it
is probable that H$_2$CO will have a similar effect on the ice
structure as CH$_3$OH and H$_2$O and thus also on the HCOOH infrared
bands. In the future both reaction paths can be tested experimentally
and thus resolve whether HCOOH is formed during ice formation, ice
processing at low temperatures or potentially even both.

\section{Summary and conclusions}
\label{summary}
The main conclusions derived from this work are:

\begin{itemize}

\item The peak positions, FWHM and profiles of the $\nu_{\rm S}$(C=O),
$\nu_{\rm B}$(OH/CH) and $\nu_{\rm S}$(C--O) bands are strongly
affected by the ice matrix. At high temperatures and when HCOOH is
deposited at high temperatures, the $\nu_{\rm S}$(C=O), $\nu_{\rm
B}$(OH/CH) and $\nu_{\rm S}$(C--O) modes split into two bands. In
binary mixtures with H$_2$O, the HCOOH features have very simple
profiles, whereas multiple bands are found in binary ice mixtures with
CO, CO$_2$ and CH$_3$OH. The peak positions of the HCOOH $\nu_{\rm
S}$(C=O), $\nu_{\rm B}$(OH/CH), and $\nu_{\rm S}$(C--O) bands are
shifted up to $\sim$40~cm$^{-1}$. The spectra for tertiary mixtures
resemble the spectra of binary mixtures with H$_2$O, but small shifts
exist.

\item The band strengths of the $\nu_{\rm S}$(C=O), $\nu_{\rm
B}$(OH/CH), and $\nu_{\rm S}$(C--O) features are strongly affected by
the ice structure. They are enhanced in mixtures of HCOOH with H$_2$O
and CO by factors of 2--10, but are similar to those of pure HCOOH ice
for binary mixtures with CO$_2$ and CH$_3$OH. Strong temperature
effects are also observed.

\item Comparison of our data with the Spitzer spectrum of HH~46 and
the ISO spectrum toward the high mass YSO W~33A excludes the presence
of the high temperature HCOOH chain-structure, as well as binary
mixtures of HCOOH with any of the other species H$_2$O, CO, CO$_2$ and
CH$_3$OH. The 7.2~$\mu$m feature toward W~33A is very similar to that
found in tertiary mixtures of $\sim$7:67:26\% HCOOH:H$_2$O:X, where X
stands for CO, CO$_2$ or CH$_3$OH. Especially the
HCOOH:H$_2$O:CH$_3$OH mixture reproduces the observed W~33A spectrum
very well. Potentially, such tertiary or even more complex ices
consisting of HCOOH, H$_2$O, CH$_3$OH, CO and CO$_2$ may be
responsible for the observed spectrum. The presence of HCOOH in such
complex ice mixtures is consistent with its possible formation
mechanisms.

\end{itemize}

\begin{acknowledgements}
  We thank the referee for useful comments on the paper. Funding was
  provided by NOVA, the Netherlands Research School for Astronomy, and
  by a Spinoza grant from the Netherlands Organization for Scientific
  Research, NWO.
\end{acknowledgements}

\Online
\begin{appendix}
\section{Line parameters of HCOOH in the 1800-1000~cm$^{-1}$ range}
\begin{table*}
\caption{Peak positions, FWHMs and $A/A_0$ for the HCOOH $\nu_{\rm
S}$(C=O), $\nu_{\rm B}$(OH/CH), and $\nu_{\rm S}$(C--O) bands for all
temperatures for pure HCOOH. The peak positions are given in both
wavenumbers and wavelengths for the ice deposited at 15~K and
145~K. The uncertainties on the peak positions and FWHMs are $\pm$1
and $\pm$2~cm$^{-1}$, respectively.}\label{pp_pure}
\begin{center}
\begin{tabular}{l|lllll|lll|lll}
\hline
\hline
             & \multicolumn{5}{c|}{$\nu_{\rm S}$(C=O)} & \multicolumn{3}{c|}{$\nu_{\rm B}$(OH/CH)} & \multicolumn{3}{c}{$\nu_{\rm B}$(C--O)}\\
             & \multicolumn{2}{c}{out-of-phase} &\multicolumn{3}{c|}{in-phase} & & & & & &\\
 $T$         & $\nu_{\rm S}$ & FWHM & $\nu_{\rm S}$ & FWHM & $A/A_{\rm 0}^{a}$ & $\nu_{\rm S}$ & FWHM & $A/A_{\rm 0^{a}}$ & $\nu_{\rm S}$ & FWHM & $A/A_{\rm 0}^{a}$ \\
 (K)           & (cm$^{-1}$($\mu$m)) & (cm$^{-1}$) & (cm$^{-1}$($\mu$m)) & (cm$^{-1}$) & &(cm$^{-1}$($\mu$m)) & (cm$^{-1}$) & & (cm$^{-1}$($\mu$m)) & (cm$^{-1}$)& \\
\hline
 & \multicolumn{11}{c}{HCOOH deposited at 15~K}\\
\hline
15           &  1714(5.84) & 45 & 1650(6.06) & 69 & 1.0 & 1387(7.21) & 37 & 1.0 & 1211(8.26) & 64 & 1.0\\ 
30           &  1713(5.84) & 43 & 1651(6.06) & 71 & 1.0 & 1387(7.21) & 37 & 1.1 & 1213(8.24) & 61 & 1.0\\ 
60           &  1710(5.85) & 41 & 1644(6.08) & 65 & 1.1 & 1387(7.21) & 35 & 1.2 & 1219(8.20) & 53 & 1.0\\ 
75           &  1710(5.85) & 40 & 1641(6.09) & 65 & 1.1 & 1387(7.21) & 33 & 1.2 & 1221(8.19) & 51 & 1.0\\ 
90           &  1709(5.85) & 39 & 1637(6.11) & 64 & 1.1 & 1387(7.21) & 31 & 1.3 & 1223(8.18) & 50 & 1.0\\ 
105          &  1709(5.85) & 39 & 1634(6.12) & 63 & 1.1 & 1387(7.21) & 30 & 1.3 & 1224(8.17) & 50 & 1.0\\ 
120          &  1709(5.85) & 41 & 1625(6.15) & 56 & 1.1 & 1386(7.21) & 29 & 1.8 & 1228(8.15) & 51 & 1.0\\ 
135          &  1703(5.87) & 10 & 1615(6.19) & 43 & 1.1 & 1391(7.19) &  8 & 2.7 & 1255(7.97) & 22 & 0.9\\ 
             & 	1716(5.83) & 39 &            &    &     & 1375(7.27) & 14 &     & 1221(8.19) & 21 &    \\
150          &  1703(5.87) & 11 & 1615(6.19) & 40 & 1.1 & 1391(7.19) &  7 & 2.5 & 1254(7.97) & 22 & 0.9\\
             & 	1719(5.82) & 34 &            &    &     & 1374(7.28) & 13 &     & 1221(8.19) & 17 &    \\
165          & 	1704(5.87) & 11 & 1617(6.18) & 36 & --- & 1391(7.19) &  7 & --- & 1253(7.98) & 20 & ---\\
             &  1722(5.81) & 32 &            &    &     & 1373(7.28) & 12 &     & 1220(8.20) & 15 &    \\
\hline
 & \multicolumn{11}{c}{HCOOH deposited at 145~K}\\
\hline
15           & 1699(5.89) &10 & 1606(6.23) & 36 & 1.5 & 1390(7.19) & 6 & 3.7 &1256(7.96) & 18 & 1.3\\
             & 1712(5.84) &32 &            &    &     & 1371(7.29) & 9 &     &1222(8.18) & 15 &    \\
30           & 1699(5.89) & 9 & 1606(6.23) & 36 & 1.5 & 1390(7.19) & 6 & 3.7 &1256(7.96) & 18 & 1.3\\
             & 1712(5.84) &32 &            &    &     & 1371(7.29) & 9 &     &1222(8.18) & 15 &    \\
45           & 1699(5.89) & 9 & 1606(6.22) & 36 & 1.5 & 1390(7.19) & 6 & 3.7 &1255(7.97) & 18 & 1.3\\
             & 1713(5.84) &32 &            &    &     & 1371(7.29) & 9 &     &1222(8.18) & 15 &    \\
60           & 1699(5.89) & 8 & 1607(6.22) & 36 & 1.5 & 1390(7.19) & 6 & 3.6 &1255(7.97) & 18 & 1.3\\
             & 1712(5.84) &32 &            &    &     & 1371(7.29) & 9 &     &1222(8.19) & 15 &    \\
75           & 1699(5.88) & 8 & 1608(6.22) & 36 & 1.5 & 1390(7.19) & 6 & 3.6 &1255(7.97) & 18 & 1.3\\
             & 1713(5.84) &32 &            &    &     & 1371(7.29) & 9 &     &1222(8.19) & 15 &    \\
90           & 1700(5.88) & 8 & 1608(6.22) & 37 & 1.5 & 1390(7.19) & 6 & 3.5 &1254(7.97) & 18 & 1.3\\
             & 1713(5.84) &32 &            &    &     & 1371(7.29) &10 &     &1221(8.19) & 15 &    \\
105          & 1700(5.88) & 9 & 1609(6.21) & 37 & 1.5 & 1390(7.19) & 6 & 3.5 &1254(7.98) & 18 & 1.3\\
             & 1713(5.84) &32 &            &    &     & 1371(7.29) &10 &     &1221(8.19) & 15 &    \\
120          & 1700(5.88) & 9 & 1610(6.21) & 38 & 1.4 & 1390(7.19) & 6 & 3.4 &1253(7.98) & 18 & 1.3\\
             & 1713(5.84) &32 &            &    &     & 1371(7.29) &10 &     &1220(8.19) & 15 &    \\
135          & 1700(5.88) & 9 & 1611(6.21) & 39 & 1.4 & 1390(7.20) & 6 & 3.3 &1252(7.99) & 18 & 1.3\\
             & 1714(5.84) &32 &            &    &     & 1371(7.29) &11 &     &1220(8.10) & 15 &    \\
150          & 1701(5.88) & 9 & 1612(6.20) & 39 & 1.4 & 1390(7.20) & 6 & 3.2 &1252(7.99) & 19 & 1.3\\
             & 1714(5.83) &31 &            &    &     & 1371(7.29) &11 &     &1219(8.20) & 14 &    \\
165          & 1701(5.88) & 9 & 1614(6.20) & 38 & 1.3 & 1390(7.20) & 6 & 2.8 &1250(8.00) & 19 & 1.2\\
             & 1715(5.83) &31 &            &    &     & 1371(7.30) &11 &     &1219(8.20) & 14 &    \\
\hline
\end{tabular}
\end{center}

$^a$ $A_{\rm0}$ refers to the band strength in pure HCOOH at 15~K.

\end{table*}

\begin{table*}
\caption{Peak positions, FWHMs and $A/A_0$ for the HCOOH $\nu_{\rm
S}$(C=O), $\nu_{\rm B}$(OH/CH), and $\nu_{\rm S}$(C--O) bands for all
temperatures for mixtures of HCOOH with H$_2$O. The uncertainties on
the peak positions and FWHMs are $\pm$1 and $\pm$2~cm$^{-1}$,
respectively.}\label{pp_hcooh-h2o}
\begin{center}
\begin{tabular}{l|lllll|lll|lll}
\hline
\hline
             & \multicolumn{5}{c|}{$\nu_{\rm S}$(C=O)} & \multicolumn{3}{c|}{$\nu_{\rm B}$(OH/CH)} & \multicolumn{3}{c}{$\nu_{\rm S}$(C--O)}\\
             & \multicolumn{2}{c}{out-of-phase} &\multicolumn{3}{c|}{in-phase} & & & & & &\\
 $T$         & $\nu_{\rm S}$ & FWHM & $\nu_{\rm S}$ & FWHM & $A/A_{\rm 0}^a$& $\nu_{\rm S}$ & FWHM & $A/A_{\rm 0}^a$ & $\nu_{\rm S}$ & FWHM & $A/A_{\rm 0}^a$ \\
 (K)           & (cm$^{-1}$($\mu$m)) & (cm$^{-1}$) & (cm$^{-1}$($\mu$m)) & (cm$^{-1}$) & &(cm$^{-1}$($\mu$m)) & (cm$^{-1}$) & &(cm$^{-1}$($\mu$m)) & (cm$^{-1}$)& \\ 
\hline
 & \multicolumn{11}{c}{50:50\% HCOOH:H$_2$O}\\
\hline
15           & 1708(5.85) & 48 & 1658(6.03) & 103& 1.4 & 1391(7.19) & 37 & 1.1 & 1215(8.23) & 50 & 1.3\\ 
30           & 1709(5.85) & 39 & 1675(5.97) & 75 & 1.5 & 1391(7.19) & 37 & 1.2 & 1220(8.20) & 58 & 1.3\\ 
45           & 1711(5.85) & 36 & 1665(6.01) & 78 & 1.5 & 1391(7.19) & 36 & 1.3 & 1223(8.18) & 55 & 1.3\\ 
60           & 1710(5.85) & 33 & 1672(5.98) & 83 & 1.5 & 1391(7.19) & 35 & 1.3 & 1223(8.17) & 54 & 1.3\\ 
75           & 1709(5.85) & 35 & 1668(6.00) & 77 & 1.5 & 1391(7.19) & 34 & 1.4 & 1224(8.17) & 52 & 1.3\\ 
90           & 1709(5.85) & 34 & 1670(6.00) & 78 & 1.5 & 1391(7.19) & 33 & 1.5 & 1225(8.17) & 51 & 1.2\\ 
105          & 1709(5.85) & 35 & 1669(5.99) & 75 & 1.5 & 1392(7.19) & 34 & 1.7 & 1225(8.16) & 51 & 1.2\\ 
120          & 1710(5.85) & 37 & 1671(5.98) & 75 & 1.5 & 1395(7.17) & 42 & 3.1 & 1224(8.17) & 50 & 1.2\\ 
135          & 1713(5.84) & 45 & 1670(5.99) & 69 & 1.5 & 1403(7.13) & 59 & 6.4 & 1223(8.17) & 51 & 0.95\\
150          & 1707(5.86) & 10 & 1616(6.19) & 40 & 1.5 & 1391(7.19) &  7 & 3.8 & 1255(7.97) & 19 & 1.2\\ 
             & 1720(5.82) & 16 &            &    &     & 1372(7.29) & 13 &     & 1222(8.18) & 17 &    \\ 
165          & 1706(5.86) & 12 & 1619(6.18) & 34 & --- & 1391(7.19) &  6 & --- & 1253(7.98) & 18 & ---\\
             & 1723(5.80) & 32 &            &    &     & 1373(7.29) & 12 &     & 1222(8.19) & 14 &    \\ 
\hline
 & \multicolumn{11}{c}{34:66\% HCOOH:H$_2$O}\\
\hline
15           & 1707(5.86) & 44 & 1652(6.05) & 95 & 1.6 & 1391(7.19) & 35 & 1.5 & 1216(8.23) & 46 & 1.3\\ 
30           & 1706(5.86) & 38 & 1688(5.92) & 94 & 1.7 & 1391(7.19) & 35 & 1.6 & 1224(8.17) & 56 & 1.3\\ 
45           & 1707(5.86) & 36 & 1667(6.00) & 83 & 1.7 & 1391(7.19) & 33 & 1.7 & 1226(8.16) & 51 & 1.3\\ 
60           & 1706(5.86) & 37 & 1668(6.00) & 77 & 1.7 & 1391(7.19) & 32 & 1.8 & 1226(8.15) & 50 & 1.3\\ 
75           & 1706(5.86) & 36 & 1669(5.99) & 75 & 1.7 & 1391(7.19) & 32 & 1.9 & 1227(8.15) & 49 & 1.3\\ 
90           & 1706(5.86) & 36 & 1670(5.99) & 74 & 1.7 & 1391(7.19) & 32 & 2.1 & 1227(8.15) & 49 & 1.3\\ 
105          & 1706(5.86) & 36 & 1671(5.98) & 73 & 1.7 & 1393(7.18) & 34 & 2.5 & 1227(8.15) & 48 & 1.3\\ 
120          & 1708(5.85) & 39 & 1674(5.98) & 73 & 1.7 & 1399(7.15) & 46 & 4.3 & 1226(8.16) & 48 & 1.2\\ 
135          & 1714(5.84) & 47 & 1670(5.99) & 69 & 1.7 & 1406(7.11) & 62 & 7.5 & 1224(8.17) & 51 & 1.0\\ 
150          & 1711(5.84) &  4 & 1617(6.18) & 42 & 1.6 & 1391(7.19) &  7 & 4.7 & 1256(7.96) & 18 & 1.2\\ 
             & 1719(5.82) & 34 &            &    &     & 1371(7.29) & 13 &     & 1223(8.17) & 16 &    \\ 
165          & ---------  & -- & 1621(6.17) & 32 & --- & 1391(7.19) &  5 & --- & 1255(7.97) & 16 & ---\\
             & 1726(5.79) & 28 &            &    &     & 1372(7.29) & 11 &     & 1224(8.17) & 13 &    \\
\hline
 & \multicolumn{11}{c}{20:80\% HCOOH:H$_2$O}\\
\hline
15           & 1707(5.86) & 40 & 1643(6.09) & 81 & 1.8 & 1391(7.19) & 35 & 2.1 & 1215(8.23) & 44 & 1.2 \\
30           & 1704(5.87) & 43 & 1667(6.00) & 89 & 1.8 & 1390(7.19) & 33 & 2.2 & 1227(8.15) & 55 & 1.3 \\
45           & 1703(5.87) & 44 & 1660(6.02) & 77 & 1.9 & 1390(7.19) & 31 & 2.3 & 1230(8.13) & 47 & 1.3 \\
60           & 1703(5.87) & 40 & 1669(5.99) & 83 & 1.9 & 1390(7.19) & 31 & 2.5 & 1229(8.13) & 48 & 1.3 \\
75           & 1703(5.87) & 42 & 1664(6.01) & 76 & 1.9 & 1391(7.19) & 31 & 2.7 & 1229(8.13) & 47 & 1.3 \\
90           & 1703(5.87) & 39 & 1670(5.99) & 79 & 1.9 & 1391(7.19) & 32 & 2.9 & 1229(8.14) & 46 & 1.3 \\
105          & 1705(5.87) & 40 & 1668(5.99) & 75 & 1.9 & 1393(7.18) & 35 & 3.5 & 1228(8.14) & 46 & 1.2 \\
120          & 1709(5.85) & 39 & 1675(5.97) & 76 & 1.9 & 1403(7.13) & 51 & 5.6 & 1227(8.15) & 47 & 1.1 \\
135          & 1714(5.83) & 45 & 1673(5.98) & 72 & 1.9 & 1413(7.08) & 61 & 8.2 & 1225(8.16) & 49 & 0.93 \\
150          & ---------  & -- & 1619(6.17) & 51 & 1.7 & 1392(7.19) &  8 & 6.1 & 1256(7.96) & 19 & 1.0 \\
             & 1717(5.82) & 41 &            &    &     & 1372(7.29) & 17 &     & 1223(8.17) & 19 &     \\
165          & ---------- & -- &-1622(6.16) & 31 &  -- & 1392(7.18) &  4 & --- & 1255(7.97) & 15 & --- \\
             & 1730(5.78) & 27 &            &    &     & 1375(7.27) & 11 &     & 1226(8.16) & 11 &     \\
\hline
 & \multicolumn{11}{c}{9.4:91\% HCOOH:H$_2$O}\\
\hline
15           & 1708(5.86) & 38 & 1636(6.11) & 70 & 2.5 & 1391(7.19) &  35 & 4.9 & 1214(8.24) & 42 & 1.4\\
30           & 1706(5.86) & 40 & 1680(5.95) &128 & 2.5 & 1388(7.21) &  31 & 4.9 & 1233(8.11) & 59 & 1.4\\
45           & 1698(5.89) & 45 & 1662(6.02) & 94 & 2.6 & 1389(7.20) &  29 & 5.2 & 1231(8.12) & 42 & 1.4\\
60           & 1699(5.89) & 55 & 1649(6.06) & 90 & 2.6 & 1388(7.20) &  28 & 5.6 & 1231(8.12) & 46 & 1.4\\
75           & 1698(5.89) & 49 & 1655(6.04) &100 & 2.6 & 1388(7.20) &  28 & 5.8 & 1231(8.12) & 45 & 1.4\\
90           & 1700(5.88) & 52 & 1641(6.10) & 88 & 2.6 & 1389(7.20) &  30 & 6.2 & 1230(8.13) & 44 & 1.4\\
105          & 1701(5.88) & 50 & 1645(6.08) & 89 & 2.6 & 1391(7.19) &  34 & 7.2 & 1230(8.13) & 44 & 1.4\\
120          & 1706(5.86) & 47 & 1664(6.01) & 94 & 2.6 & 1402(7.13) &  52 & 9.5 & 1228(8.14) & 45 & 1.3\\
135          & 1713(5.84) & 46 & 1658(6.03) & 76 & 2.5 & 1415(7.06) &  59 &11.3 & 1226(8.15) & 47 & 1.1\\
150          & 1718(5.82) & 43 & 1674(5.97) & 93 & 2.3 & 1404(7.12) &  67 &10.5 & 1223(8.18) & 51 & 1.1\\
\hline	     		    
\end{tabular}
\end{center}

$^a$ $A_{\rm0}$ refers to the band strength in pure HCOOH at 15~K.

\end{table*}

\begin{table*}
\caption{Peak positions, FWHMs and $A/A_0$ for the HCOOH $\nu_{\rm
 S}$(C=O) and $\nu_{\rm S}$(C--O) bands for all temperatures for
 $\sim$10:90\% HCOOH:CH$_3$OH. The uncertainties on the peak positions
 and FWHMs are $\pm$1 and $\pm$2~cm$^{-1}$,
 respectively.}\label{pp_hcooh-ch3oh}
\begin{center}
\begin{tabular}{l|lllll|lllll}
\hline
\hline
             & \multicolumn{5}{c|}{$\nu_{\rm S}$(C=O)} &  \multicolumn{5}{c}{$\nu_{\rm S}$(C--O)}\\
             & \multicolumn{2}{c}{out-of-phase} &\multicolumn{3}{c|}{in-phase} & \multicolumn{2}{c}{out-of-phase} &\multicolumn{3}{c|}{in-phase}\\
 $T$         & $\nu$ & FWHM & $\nu$ & FWHM & $A/A_{\rm 0}^a$ & $\nu$ & FWHM & $\nu$ & FWHM & $A/A_{\rm 0}^a$\\
 (K)           & (cm$^{-1}$($\mu$m)) & (cm$^{-1}$) & (cm$^{-1}$($\mu$m)) & (cm$^{-1}$) & & (cm$^{-1}$($\mu$m)) & (cm$^{-1}$)& (cm$^{-1}$($\mu$m)) & (cm$^{-1}$)& \\ 
\hline
15           & 1721(5.81) & 15 & 1691(5.91) & 31 & 1.2 & 1231(8.12) & 29 & 1198(8.35) & 26 & 0.86\\  
45           & 1717(5.82) & 18 & 1690(5.92) & 34 & 1.2 & 1235(8.10) & 24 & 1203(8.31) & 21 & 0.85\\  
75           & 1716(5.83) & 11 & 1689(5.92) & 33 & 1.2 & 1234(8.10) & 24 & 1201(8.33) & 18 & 0.85\\  
105          & 1716(5.83) & 13 & 1686(5.93) & 28 & 1.2 & 1234(8.10) & 21 & 1202(8.32) & 18 & 0.86\\  
135          & 1725(5.80) & 13 & 1696(5.90) & 44 & 1.2 & 1230(8.13) & 27 & 1201(8.33) & 22 & 0.63\\  
\hline
\end{tabular}
\end{center}

$^a$ $A_{\rm0}$ refers to the band strength in pure HCOOH at 15~K.
\end{table*}

\begin{table*}
\caption{Peak positions and band strengths for the $\nu_{\rm S}$(C=O)
and $\nu_{\rm S}$(C--O) bands for the binary $\sim$10:90\% HCOOH:CO
and HCOOH:CO$_2$ mixtures and tertiary $\sim$7:67:26\%
HCOOH:H$_2$O:CO, HCOOH:H$_2$O:CO$_2$ and HCOOH:H$_2$O:CH$_3$OH
mixtures at 15~K. The uncertainties on the peak positions are
$\pm$1~cm$^{-1}$.}\label{nuc=o}
\begin{tabular}{l|ll}
\hline
\hline
Mode & $\nu$  & $A/A_{\rm 0}^a$\\
     & (cm$^{-1}$)  & \\
\hline 
\multicolumn{3}{c}{11:89\% HCOOH:CO}\\
\hline
$\nu_{\rm S}$(C=O) & 1743,1735, 1721, 1702 & 1.9\\
$\nu_{\rm B}$(OH/CH) & 1390              & 3.0\\
$\nu_{\rm S}$(C--O) & 1250, 1235, 1227, 1218, 1185, 1172, 1130 & 2.1\\
\hline
\multicolumn{3}{c}{9:91\% HCOOH:C$^{18}$O$_2$}\\
\hline
$\nu_{\rm S}$(C=O) & 1744, 1734, 1717, 1612 & 0.72\\
$\nu_{\rm B}$(OH/CH) & 1387 & 1.4\\
$\nu_{\rm S}$(C--O) & 1227, 1187, 1157, 1132 & 0.63\\
\hline
\multicolumn{3}{c}{8:62:30\% HCOOH:H$_2$O:CO}\\
\hline
$\nu_{\rm S}$(C=O) & 1705, 1685 & 4.7\\
$\nu_{\rm S}$(C--O) & 1227 & 2.4\\
\hline
\multicolumn{3}{c}{6:67:27\% HCOOH:H$_2$O:C$^{18}$O$_2$}\\
\hline
$\nu_{\rm S}$(C=O) & 1695 & 4.2\\
$\nu_{\rm S}$(C--O) & 1229 & 1.5\\
\hline
\multicolumn{3}{c}{6:68:26\% HCOOH:H$_2$O:CH$_3$OH}\\
\hline
$\nu_{\rm S}$(C=O) & 1703 & 3.7\\
$\nu_{\rm S}$(C--O) & 1223 & 1.5\\
\hline
\end{tabular}

$^a$ $A_{\rm0}$ refers to the band strength in pure HCOOH at 15~K.
\end{table*}

\end{appendix}

\end{document}